\documentclass[aps,floats]{revtex4}
\usepackage{amsmath,amssymb}
\usepackage{graphicx,epsfig}

\begin{document}
\bibliographystyle {plain}

\def\oppropto{\mathop{\propto}} 
\def\opmin{\mathop{\min}} 
\def\opmax{\mathop{\max}} 
\def\opsimeq{\mathop{\simeq}}
\def\opoverderline{\mathop{\overline}}
\def\operarrow{\mathop{\longrightarrow}}
\def\opsim{\mathop{\sim}}

\def\fig#1#2{\includegraphics[height=#1]{#2}}
\def\figx#1#2{\includegraphics[width=#1]{#2}}


\title{  Anderson transition on the Cayley tree \\ as a traveling wave
critical point for various probability distributions } 


 \author{ C\'ecile Monthus and Thomas Garel }
  \affiliation{Institut de Physique Th\'{e}orique, CNRS and CEA Saclay
 91191 Gif-sur-Yvette cedex, France}

\begin{abstract}
For Anderson localization on the Cayley tree, we study the statistics of various observables as a function of the disorder strength $W$ and the number $N$ of generations. We first consider the Landauer transmission $T_N$. In the localized phase, its logarithm follows the traveling wave form $\ln T_N \simeq \overline{\ln T_N} + \ln t^*$ where (i) the disorder-averaged value moves linearly $\overline{\ln (T_N)} \simeq - \frac{N}{\xi_{loc}}$ and the localization length diverges as $\xi_{loc} \sim (W-W_c)^{-\nu_{loc}}$ with $\nu_{loc}=1$ (ii) the variable $t^*$ is a fixed random variable with a power-law tail $P^*(t^*) \sim 1/(t^*)^{1+\beta(W)}$ for large $t^*$ with $0<\beta(W) \leq 1/2$, so that all integer moments of $T_N$ are governed by rare events. In the delocalized phase, the transmission $T_N$ remains a finite random variable as $N \to \infty$, and we measure near criticality the essential singularity $\overline{\ln (T_{\infty})} \sim - \vert W_c-W \vert^{-\kappa_T}$ with $\kappa_T \sim 0.25$. We then consider the statistical properties of normalized eigenstates $\sum_x \vert \psi(x) \vert^{2}=1$, in particular the entropy $S = - \sum_x \vert \psi(x) \vert^{2} \ln \vert \psi(x) \vert^{2}$ and the Inverse Participation Ratios (I.P.R.) $I_q = \sum_x \vert \psi(x) \vert^{2q}$. In the localized phase, the typical entropy diverges as $S_{typ}  \sim  ( W-W_c)^{- \nu_S}$ with $\nu_S \sim 1.5$, whereas it grows linearly $S_{typ}(N) \sim  N$ in the delocalized phase. Finally for the I.P.R., we explain how closely related variables propagate as traveling waves in the delocalized phase. In conclusion, both the localized phase and the delocalized phase are characterized by the traveling wave propagation of some probability distributions, and the Anderson localization/delocalization transition then corresponds to a traveling/non-traveling critical point. Moreover, our results point towards the existence of several length scales that diverge with different exponents $\nu$ at criticality.

\end{abstract}

\maketitle

\section{ Introduction }

Since its discovery fifty years ago \cite{anderson}
Anderson localization has remained a very active field
of research (see for instance the reviews
 \cite{thouless,souillard,bookpastur,Kramer,markos,mirlinrevue}).
According to the scaling theory \cite{scaltheo}, 
there is no delocalized phase in dimensions $d=1,2$,
whereas there exists a localization/delocalization at finite disorder
in dimension $d>2$. To get some insight into this type of transition,
it is natural to consider Anderson localization on the Cayley tree
 (or Bethe lattice)
which is expected to represent some mean-field limit.
The tight-binding Anderson model on the Cayley
 tree has been thus studied by various
authors. 
In \cite{abou}, the recursion equation for the self-energy has been studied
to establish the mobility edge as the limit of stability of localized states.
In \cite{Kun_Sou}, 
the exponent $\nu$ governing the divergence of the localization
length was shown to be $\nu=1$. 
In \cite{MirlinBethe}, the supersymmetric formalism
has been used to predict the critical behavior of some
disorder-averaged observables 
 (see also \cite{zirnbauer,verbaar,efetov2lengths,efetovbook}
 where similar results have been obtained for the case where
 Anderson tight-binding model is replaced by a non-linear $\sigma$-model).
In \cite{DR,MD}, the problem was reformulated in terms of recursions on 
Riccati variables to obtain weak-disorder expansions.
Other studies have focused on random-scattering models on the Cayley tree
\cite{shapiro,chalker,bell}. 
More recently, the interest in Anderson localization on the Cayley tree
has been revived by the question of many-body localization \cite{manybodyloc},
 because the geometry
of the Fock space of many-body states was argued to be similar to a Cayley
 tree \cite{levitov,silvestrov,gornyi,huse}.

	In the present paper, we consider the Anderson tight binding model on
the Cayley tree already studied in \cite{abou,Kun_Sou,MirlinBethe,DR,MD}
and we study numerically the statistical properties of Landauer transmission and eigenstates
as a function of the disorder strength $W$
 and of the number $N$ of generations.
We find that several probability distributions 
propagate as traveling waves with a fixed shape,
 as the number $N$ of tree generations grows.
The fact that traveling waves appear in disordered models defined on trees
has been discovered by Derrida and Spohn \cite{Der_Spohn}
on the specific example of the directed polymer in a random medium 
 and was then found in various models \cite{majumdar}.
For the case of Anderson localization on trees,
the analysis of \cite{abou} concerning the distribution of the self-energy
in the localized phase is actually a 'traveling wave analysis' 
(see Appendix A for more details),
 although it is not explicitly mentioned in these terms in \cite{abou}.
The fact that traveling waves occur has been explicitly seen
in the numerical study of the transmission distribution for
a random-scattering model on the Cayley tree (see Fig. 8b of \cite{bell})
and has been found within the supersymmetric formalism
\cite{zirnbauer,efetov2lengths,efetovbook}.
In the field of traveling waves and front propagation 
(see the reviews \cite{revues_traveling,brunetreview}), 
there exists an essential separation between two classes : 
in 'pulled fronts', the velocity is determined
 by the form of the tail of the front
and thus by the appropriate linearized equation in the tail region,
whereas in 'pushed fronts', the velocity is determined by the bulk properties
and thus by the non-linear dynamics in the bulk region. 
It turns out that for disordered systems defined on trees, 
the traveling waves that appear usually correspond
 to cases where it is the tail of the 
probability distribution that determines the velocity \cite{Der_Spohn}. 
In particular for Anderson localization on the Cayley tree, 
the traveling wave propagation
is also determined by the tails \cite{abou}.
From this traveling wave point of view, the localization/delocalization
Anderson transition thus represents a traveling/non-traveling critical point.
Such a traveling/non-traveling phase transition 
for a branching random walk in the presence of a moving absorbing wall
 has been studied recently in \cite{simon} and we find here very similar 
behaviors in the critical region.

The paper is organized as follows.
In Section \ref{trans}, we describe the statistical properties
of Landauer transmission. In Section \ref{eigen}, we discuss
the statistical properties of eigenstates, as measured by
the entropy and Inverse Participation Ratios (I.P.R.).
We summarize our conclusions in Section \ref{conclusion}.
In Appendix A, we translate the analysis of Ref. \cite{abou}
concerning the distribution of the self-energy in the localized phase
into a traveling-wave tail analysis for the Landauer transmission
discussed in section \ref{trans}. In Appendix B, we explain how such 
a similar tail analysis can be performed in the delocalized phase for auxiliary
variables that are closely related to Inverse Participation Ratios.
Finally, in Appendix C, we recall the results of Ref. \cite{simon}
concerning the traveling/non-traveling phase transition 
for a branching random walk in the presence of a moving absorbing wall,
since these results are used as a comparison in the text
 to understand the finite-size scaling properties
in the critical region.

\section{  Statistical properties of the Landauer transmission }

\label{trans}

In quantum coherent problems, the most appropriate characterisation 
of transport properties consists in defining a scattering problem
where the disordered sample is linked to incoming wires and outgoing wires
and in studying the reflexion and transmission coefficients.
This scattering theory definition of transport, first introduced by Landauer \cite{landauer},
has been much used for one-dimensional systems \cite{anderson_fisher,anderson_lee,luck}
and has been generalized to higher dimensionalities and multi-probes
measurements (see the review \cite{stone}).
In dimension $d=1$, the transfer matrix formulation of the Schr\"odinger 
equation yields that the probability distribution of the Landauer transmission
becomes asymptotically log-normal \cite{anderson_fisher,luck} , i.e.  one has
\begin{eqnarray}
\ln T_L^{(1d)} = - \frac{L}{\xi_{loc}} + L^{1/2} u
\label{trans1d}
\end{eqnarray}
where $\xi_{loc} $ represents the localization length and where $u$ is a sample-dependent random variable of order $O(1)$
distributed with a Gaussian law.
Although it is often assumed that this log-normal distribution 
persists in the localized phase in dimension $d=2,3$, recent numerical studies \cite{prior}
are in favor of the following scaling form for the logarithm of the conductance
\begin{eqnarray}
\ln g_L^{(d)} = - \frac{L}{\xi_{loc}} + L^{\omega(d)} u
\label{transd}
\end{eqnarray}
with exponents of order $\omega(d=2) \simeq 1/3$ and $\omega(d=3) \simeq 1/5$ \cite{prior},
whereas $\omega(d=1)=1/2$ from Eq. \ref{trans1d}.
For the Cayley tree that we consider in this paper, we will find below that the fluctuation exponent vanishes $\omega_{Cayley}=0$,
and we will discuss the probability distribution of the variable $u$.
But let us first recall the appropriate scattering framework for the Cayley tree  \cite{MD} .

\subsection{ Reminder on the Miller-Derrida framework to compute the Landauer
transmission \cite{MD}} 

We consider the Anderson tight-binding model
\begin{eqnarray}
H = \sum_i \epsilon_i \vert i > < i \vert  +  \sum_{<i,j>}  \vert i > < j \vert
\label{handerson}
\end{eqnarray}
where the hopping between nearest neighbors $<i,j>$ is a constant $V=1$ 
and where the on-site energies $\epsilon_i$ are independent random variables
drawn from the flat distribution 
\begin{eqnarray}
p(\epsilon_i) = \frac{1}{W}
 \theta \left( - \frac{W}{2} \leq \epsilon_i \leq \frac{W}{2}  \right)
\label{flat}
\end{eqnarray}
The parameter $W$ thus represents the disorder strength.

\begin{figure}[htbp]
 \includegraphics[height=6cm]{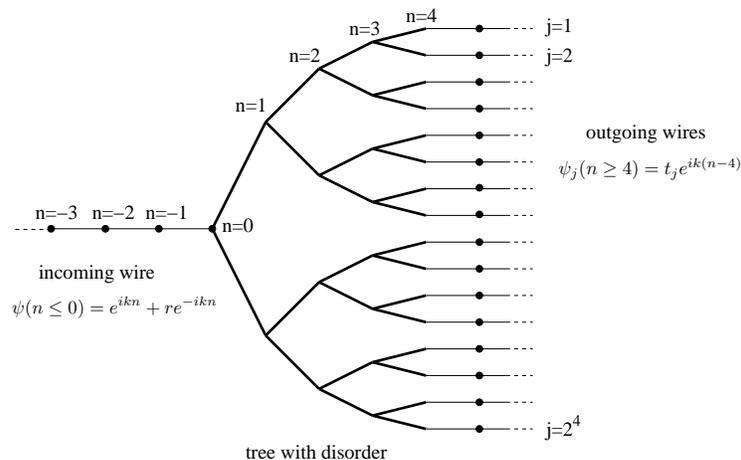}
\caption{ Scattering geometry of Ref. \cite{MD} : the disordered tree 
of branching ratio $K=2$ starting at generation $n=0$
 and ending at generation $2N$ (on the Figure $2N=4$)
is attached to one incoming wire and to $K^{2N}$ outgoing wires. 
In section \ref{trans}, we discuss the properties of the 
total transmission
$T \equiv \sum_j \vert t_j \vert^2 = 1 - \vert r \vert^2 $
where $r$ is the reflexion amplitude of the incoming wire, 
and $t_j$ the transmission
amplitudes of the outgoing wires.}
\label{figscattering}
\end{figure}

We consider the scattering geometry introduced in \cite{MD}
and shown on Fig. \ref{figscattering} :
 the finite tree of branching ratio $K$ 
is attached to one incoming wire at its root (generation $n=0$)
and to $K^{2N}$ outgoing wires at generation $2N$.
One is interested into the eigenstate $\vert \psi >$ that satisfies 
the  Schr\"odinger equation
\begin{eqnarray}
H \vert \psi > = E \psi > 
\label{schodinger}
\end{eqnarray}
inside the disorder sample and in the wires
 where one requires the plane-wave forms
\begin{eqnarray}
\psi(n \leq 0) && = e^{ik n} +r e^{- i k n} \nonumber \\
 \psi_j(n \geq 2 N) && = t_j e^{ik (n-2 N)} 
\label{psiqoutside}
\end{eqnarray}
These boundary conditions define
 the reflexion amplitude $r$ of the incoming wire
and the transmission amplitudes $t_j$ of the $j=1,2,..K^{2N}$ outgoing wires.
To satisfy the Schr\"odinger Equation of Eq. \ref{schodinger} within the wires with
the forms of Eq. \ref{psiqoutside}, one has the following relation between
the energy $E$ and the wave vector $k$  
\begin{eqnarray}
 E=2 \cos k  
\label{relationEk}
\end{eqnarray}
To simplify the discussion, we will focus in this paper on the case of
zero-energy $E=0$ and wave-vector $k=\pi/2$
\begin{eqnarray}
{ \rm In \ \ this \ \ paper :   }  \ \ E=0  \ \  { \rm and  }  \ \ k=\pi/2
\label{zeroenergy}
\end{eqnarray}
because the zero-energy $E=0$ corresponds to the center of the band 
where the delocalizion first appears when the strength $W$ of the disorder
is decreased from the strong disorder localized phase.

Inside the Cayley tree $0 \leq n \leq 2N-1$,
the Schr\"odinger Equation of Eq. \ref{schodinger} involves
 one ancestor denoted by $anc(n,j)$ and $K$ descendants denoted by $des_m(n,j)$
\begin{eqnarray}
0 = \epsilon(n,j) \psi(n,j) + \psi(anc(n,j))  + \sum_{m=1}^K \psi(des_m(n,j))
\label{schodingerinside}
\end{eqnarray}
whereas for the last generation $2N$ there is only one ancestor
 and one descendant (outgoing wire)
\begin{eqnarray}
0 = \epsilon(2N,j) \psi(2N,j) + \psi(anc(2N,j)) + \psi(2N+1,j)
\label{schodingerbord}
\end{eqnarray}

As explained in \cite{MD}, it is convenient to introduce the Riccati variables
\begin{eqnarray}
R(n,j) \equiv \frac{ \psi(anc(n,j))}{\psi(n,j) }
\label{riccati}
\end{eqnarray}
that represent the ratio of the wave function of two neighboring sites.
On the outgoing wires, these Riccati variables are fixed by Eq. \ref{psiqoutside}  to be
\begin{eqnarray}
R(2N+1,j)= \frac{\psi(2N,j)}{\psi(2N+1,j)} = e^{-ik}= -i
\label{riccatioutgoing}
\end{eqnarray}
The Schr\"odinger equation of Eq. \ref{schodingerbord} gives the 
first recursion 
\begin{eqnarray}
R(2N,j) = - \epsilon(2N,j)  - \frac{1}{R(2N+1,j)} = - \epsilon(2N,j) - e^{ik}
= - \epsilon(2N,j) -i
\label{riccatifirst}
\end{eqnarray}
whereas the Schr\"odinger equation of Eq. \ref{schodingerinside} gives the 
 recursion inside the tree for $0 \leq n \leq 2N-1$
\begin{eqnarray}
R (n,j) = - \epsilon(n,j) - \sum_{m=1}^K \frac{1}{ R(des_m(n,j)) }
\label{riccatiinside}
\end{eqnarray}
On the other hand, the value of the Riccati variable for 
the origin of the tree $n=0$ is fixed by the incoming wire of Eq. \ref{psiqoutside}
\begin{eqnarray}
R (0) = \frac{ \psi(-1) } { \psi(0) } = \frac{ e^{-ik} + r e^{ik}}{1+r}
\label{riccatiincoming}
\end{eqnarray}
and thus the reflexion coefficient $r$ can be obtained via 
\begin{eqnarray}
r= - \frac{ R(0) - e^{-ik}}{R(0)- e^{ik}} =  \frac{i+R(0)}{i-R(0)}
\label{reflexioncoef}
\end{eqnarray}
from the $R(0)$ obtained via the recursion of Eq. \ref{riccatiinside}.
From the conservation of energy, the total transmission $T$ is related to the 
reflexion coefficient $\vert r \vert^2$
\begin{eqnarray}
T \equiv \sum_j \vert t_j \vert^2 = 1 - \vert r \vert^2
\label{deftotaltrans}
\end{eqnarray}

As explained in \cite{MD}, the criterion for the localization/delocalization 
phases is then the following : 

(a) if the Riccati variable $R(0)$ at the root of the tree
converges towards a real random variable as $N \to \infty$,
the reflexion is total $\vert r \vert =1$ and the transmission vanishes $T=0$.

(b) if the Riccati variable $R(0)$ keeps a finite negative
 imaginary part as $N \to \infty$,
the reflexion is only partial $\vert r \vert <1$ and 
the transmission remains finite $T = 1 - \vert r \vert^2 >0$

We refer to \cite{MD} for the results of
a weak disorder expansion within this framework,
and for a numerical Monte-Carlo approach to determine the mobility edge
in the plane $(E,W)$.
Here we will instead study the statistical properties
of the transmission $T$ at zero energy $E=0$ as a function of the disorder
strength $W$ (Eq \ref{flat}) and of the number $N$ of generations.
 But before discussing the disordered case,
let us first describe the finite-size properties of the pure case.

\subsection{ Example : transmission of finite pure trees  }

\label{puretrans}

In the pure case where all on-site energies vanish $\epsilon(n,j)=0$,
 all branches are equivalent and there is no dependence on $j$. 
The recursions for Riccati variables of Eqs \ref{riccatioutgoing},
 \ref{riccatifirst}
and \ref{riccatiinside} give
\begin{eqnarray}
R_{pure}(2N+1) && = -i \nonumber \\
R_{pure}(2N) && = -i \nonumber \\
R_{pure} (n) && =  - \frac{K}{ R_{pure}(n+1) } \ \ {\rm for} \ \ 0 \leq n \leq 2N-1
\label{riccatipure}
\end{eqnarray}
i.e. one obtains the simple alternation between odd and even generations 
\begin{eqnarray}
R_{pure} (2N-1) && =  - \frac{K}{ R_{pure}(2N) } = -K i \nonumber \\
R_{pure} (2N-2) && =  - \frac{K}{ R_{pure}(2N-1) } = -  i \nonumber \\
... && ... \\
R_{pure} (1) && =  -K i \nonumber \\
R_{pure} (0) && =  -  i \nonumber \\
\label{riccatiinsidepuredetails}
\end{eqnarray}
The reflexion coefficient $r$ thus vanishes exactly for any even size $(2N)$
\begin{eqnarray}
r_{pure}=   \frac{i+R_{pure}(0)}{i-R_{pure}(0)} = 0
\label{reflexioncoefpure}
\end{eqnarray}
and the transmission coefficient in each branch $j=1,..,K^{2N}$ reads
\begin{eqnarray}
t_j^{pure} = (1+r)  \prod_{n=1}^{2N} \frac{1}{R_{pure}(n)} = \frac{1}{(-K)^N}
\label{tjpure}
\end{eqnarray}
(Note that for pure trees of uneven size, the reflexion would not vanish.
This is why in the disordered case, we will only consider trees of even sizes $2N$
where in the corresponding pure case, the reflexion vanish exactly for any finite tree).
It is instructive to write now the corresponding wave function
as a function of the generation n,
\begin{eqnarray}
... \ \ \ 
\psi_{pure}(-4) && =1 \nonumber \\
\psi_{pure}(-3) && =i \nonumber \\
\psi_{pure}(-2) && =-1 \nonumber \\
\psi_{pure}(-1) && =-i \nonumber \\
\psi_{pure}(0) && =1 \nonumber \\
\psi_{pure}(1) && = \frac{i}{K} \nonumber \\
\psi_{pure}(2) && = -\frac{1}{K} \nonumber \\
\psi_{pure}(3) && = - \frac{i}{K^2} \nonumber \\
\psi_{pure}(4) && =  \frac{1}{K^2} \ \ \ ...
\label{psipure}
\end{eqnarray}
This exponential decay of the wavefunction in this delocalized case
is very peculiar to the tree geometry : 
it is imposed by the energy conservation
and by the exponential growth of the number of sites with the generation $n$.

\subsection{ Statistics over the disordered samples of the transmission $T_N$}

\label{totaltrans}

\subsubsection{ Numerical pool method } 

\label{poolmethod}

If one wishes to study numerically real trees, one is limited to rather small 
number of generations of order $N_{max} \sim 12,14$
(see for instance the study \cite{Sade_Ber} based on exact diagonalization) because the number of sites and thus of
random energies grows exponentially in $N$.
From the point of view of convergence towards stable probability
distributions via recursion relations, it is thus better
 to use the so-called 'pool method' 
that will allow us to study much larger number of generations
 (see below Eq. \ref{pool5}).
The idea of the pool method is the following :  at each generation, one keeps the same number $M_{pool}$
of random variables to represent probability distributions.
Within our present framework, the probability distribution $P_n(R)$ of the 
Riccati variables $R$ at generation $n$
 will be represented by a pool of $M_{pool}$ complex values 
$\{R_n^{(1)},..,R_n^{(M_{pool})} \}$. 
It is convenient from now on to change the notation $n \to 2N-n$ with respect to Fig. \ref{figscattering}
so that $n=0$ now corresponds to the contacts with the outgoing wires (see Fig. \ref{figscattering}).
The numerical results presented below have been obtained with the following procedure :

(i) the random variables of the initial pool
 are given by (Eq. \ref{riccatifirst})
\begin{eqnarray}
R_0(j) =  - \epsilon_0(j) -i
\label{poolzero}
\end{eqnarray}
where $\epsilon(j)$ are independent random variables drawn with Eq. \ref{flat}.

(ii) From the random variables $R_{n-1}(j)$ of the pool at generation $n-1$,
the pool at generation $n$ is constructed as follows.
For each $j=1,2,..M_{pool}$, one generates a new random energy $\epsilon_n(j)$
with the law of Eq. \ref{flat} and one draws $K$ random indices 
$\{j_1(j),..j_K(j) \}$ among the pool of generation $(n-1)$ to 
construct the variable $R_{n}(j)$ as follows(Eq. \ref{riccatiinside})
\begin{eqnarray}
R_n (j) = - \epsilon_n(j) - \sum_{m=1}^K \frac{1}{ R_{n-1}(j_m(j)) }
\label{riccatiinsideresume}
\end{eqnarray}

(iii)
From these pools of Riccati variables, 
one may compute via Eqs \ref{reflexioncoef}
and \ref{deftotaltrans}
a pool of total transmission $T_n(j)$ for trees of $n$ generations using  
\begin{eqnarray}
T_n(j)= 1- \vert r_n(j) \vert^2 = 
1 - \left \vert \frac{i+R_n(j)}{i-R_n(j)} \right \vert^2
\label{transresume}
\end{eqnarray}

For the Anderson model on the Bethe lattice, 
the pool method has been already used, in particular 
in \cite{abou} with a pool $M_{pool} = 1800$
 with a number $N_{max} \sim 30$ of generations,
in \cite{MD} with pools up to $M_{pool} = 10 000$,
and in \cite{bell} with pools up to $M_{pool} =16384 $ with 
a number $N_{max} \sim 400$ of generations.
The pool method is also very much used for
 disordered systems on hierarchical lattices
(see for instance\cite{poolspinglass,Coo_Der,diamondpolymer}),

In the remaining of this section, we present the numerical results obtained 
with a pool of size
\begin{eqnarray}
M_{pool}=10^5 \ \ \ {\rm with \ \ a \ \ number \ \ of \ \ generations }
 \ \ N \leq N_{max}=34.10^5
\label{pool5}
\end{eqnarray}
We have also results for a pool of size $M_{pool}=10^6 $ with  a number of  generations $N \leq N_{max}=24.10^4$
to see how the results change with the pool size. However, 
the number of generations $N_g$ for this bigger pool $M_{pool}=10^6$
 has turned out to be less precise in the critical region. 
All figures shown below thus corresponds
 to data obtained with the pool of size $M_{pool}=10^5$.

As is usual with the pool method \cite{diamondpolymer}, 
the location of the critical point
depends on the pool, i.e. on the discrete sampling with $M_{pool}$ values of 
probability distributions. It is expected to converge towards the thermodynamic
critical point only in the limit $M_{pool} \to \infty$ 
(see the discussion of section \ref{fspool} in Appendix B).
Nevertheless, for each given pool, the critical behaviors with respect to this
pool-dependent critical point usually
 allows a good measure of critical exponents \cite{diamondpolymer}.
For instance, for the pool of size  $M_{pool}=10^5$, the critical value $W_c$
of the disorder strength where the localization-delocalization occurs 
for the Landauer transmission is of order
\begin{eqnarray}
W_c (M_{pool}=10^5 ) \simeq 16.99..
\label{critipoole5}
\end{eqnarray}
For the pool of size $M_{pool}=10^6$, we find that it is higher and of
 order $W_c (M_{pool}=10^6) \simeq 17.32..$.
This rather important shift of the pseudo-critical point
 with the pool size which has already been seen in  \cite{MD},
 seems to be due to the very slow logarithmic convergence of traveling wave velocity in the presence of cut-off
 (see the discussion of section \ref{fspool} in Appendix B). 
So we stress that here, in contrast to \cite{abou,MD},
 our goal is not to determine the true thermodynamic mobility edge 
$W_c (+\infty )$, but instead to understand the critical behaviors
 of the finite-pool results with respect
 to the pool-dependent critical point of Eq. \ref{critipoole5}.

We will first discuss the behavior of 
 the typical transmission $T_N^{typ}$ defined by 
\begin{eqnarray}
\ln (T_N^{typ}) \equiv \overline{ \ln T_N }
\label{defTtyp}
\end{eqnarray}
as a function of the number $N$ of generations and disorder strength $W$,
before we turn to the distribution around this typical value.

\subsubsection{ Exponential decay of the typical transmission $T_N^{typ} $
in the localized phase $W>W_c$ }

\begin{figure}[htbp]
 \includegraphics[height=6cm]{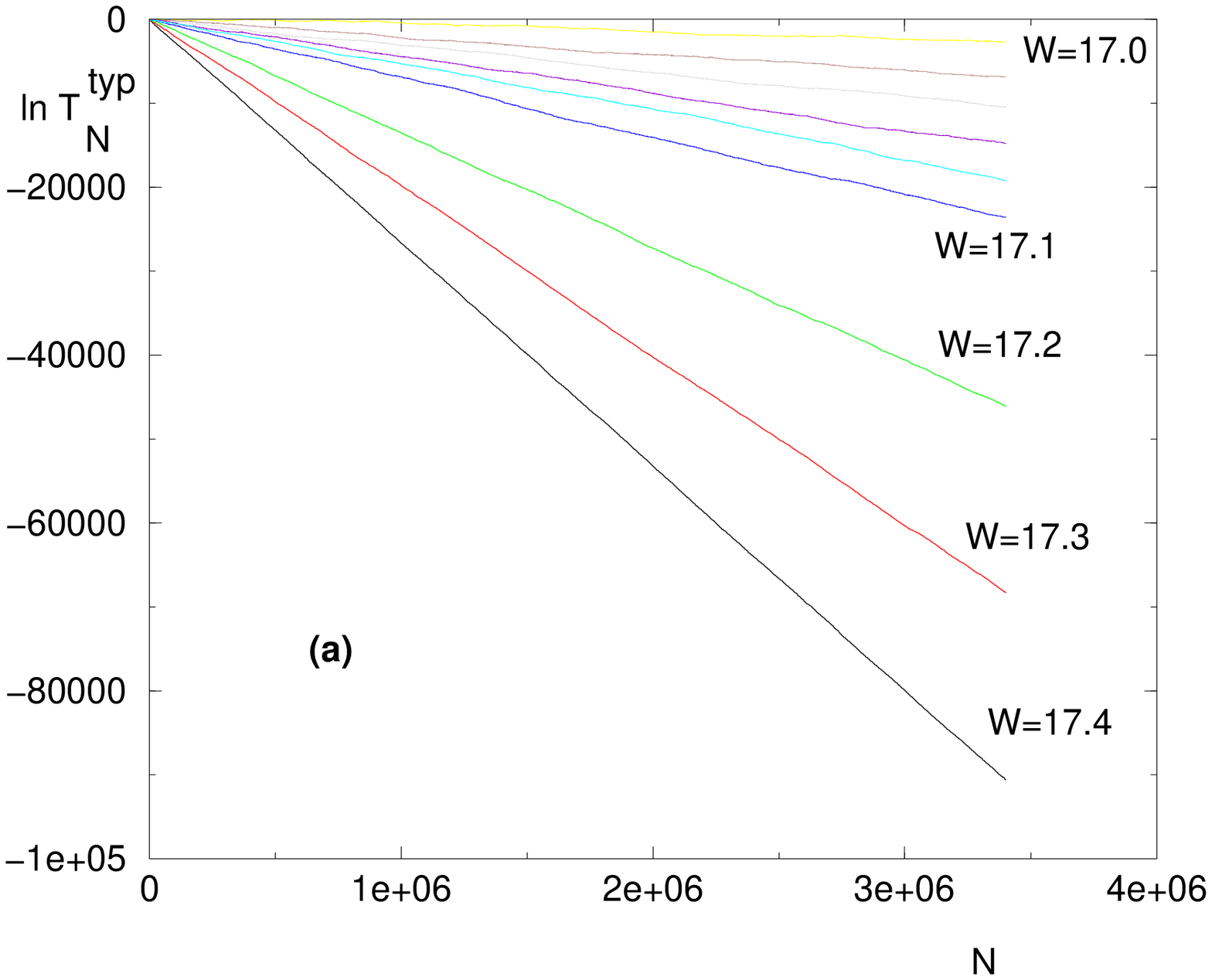}
\hspace{2cm}
\includegraphics[height=6cm]{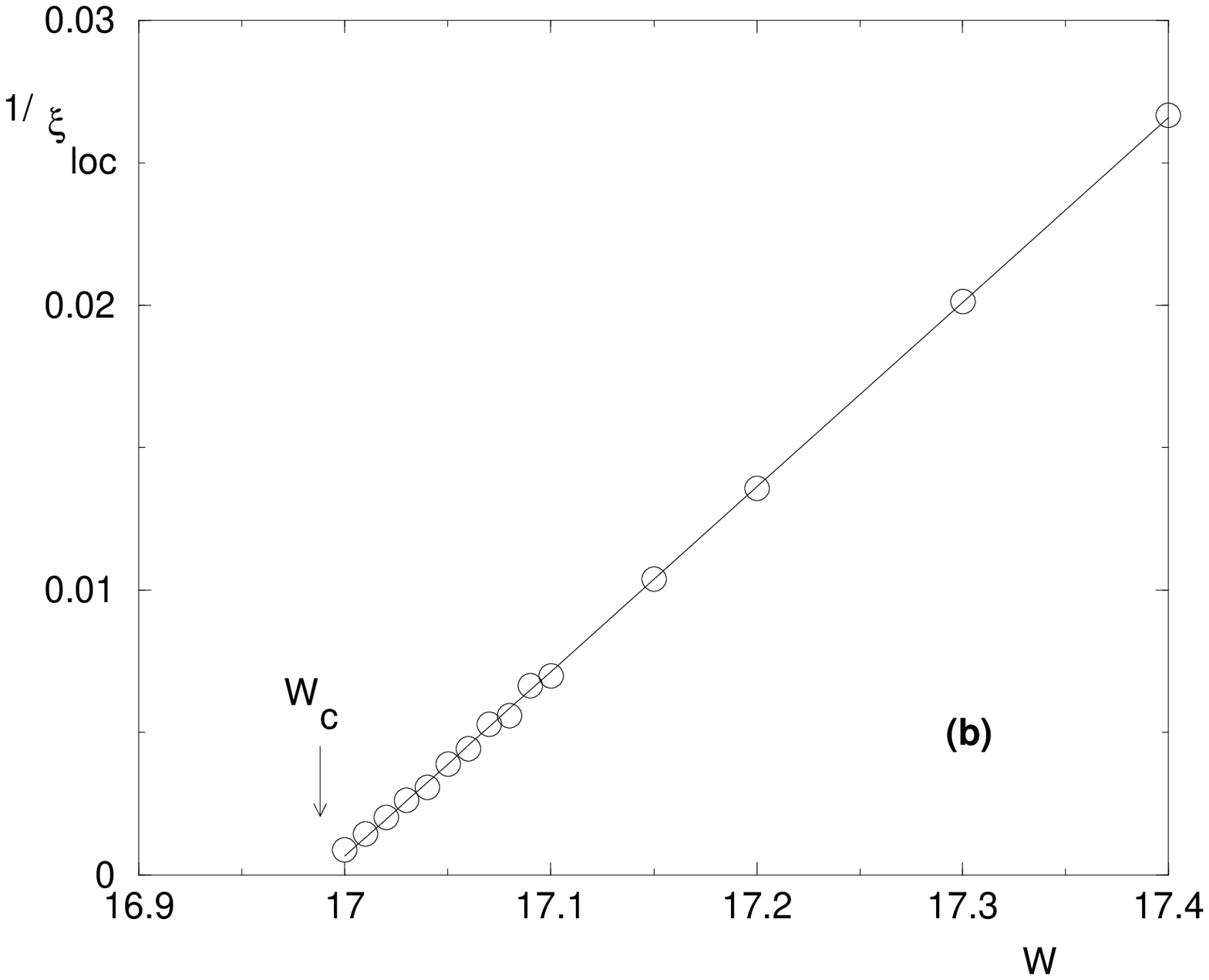}
\caption{ Exponential decay of the typical transmission 
$T_N^{typ} \equiv e^{\overline{ \ln T_N }}$ in the localized phase $W>W_c$ :
(a) Linear decay of $\overline{ \ln T_N }$ 
as a function of the number $N$ of generations (see Eq. \ref{transloc}).
(b) Behavior of the slope $1/\xi_{loc}(W)$ as a function
 of the disorder strength $W$ :
it vanishes linearly $1/\xi_{loc}(W) \sim (W-W_c^{pool})^{\nu_{loc}}$  
with $W_c^{pool} \simeq 16.99$ (see Eq. \ref{critipoole5} ) and
$\nu_{loc}=1$ (see Eq. \ref{xiloc}) }
\label{figtransloc}
\end{figure}

In the localized phase, one expects that the typical transmission $T_N^{typ}$
defined by Eq. \ref{defTtyp} decays exponentially
 with the number $N$ of generations
\begin{eqnarray}
\ln (T_N^{typ}) \equiv \overline{ \ln T_N(W>W_c) } \opsimeq_{N \to \infty} - \frac{N}{\xi_{loc}(W)}
\label{transloc}
\end{eqnarray}
where $\xi_{loc}$ represents the localization length that diverges
at the delocalization transition
\begin{eqnarray}
\xi_{loc}(W) \opsimeq_{W \to W_c^+} (W-W_c)^{-\nu_{loc}}
\label{xiloc}
\end{eqnarray}.

We show on Fig. \ref{figtransloc} our numerical results
 for the pool of size $M_{pool}=10^5$ (Eq. \ref{pool5}) : 
the exponential decay with $N$ of Eq. \ref{transloc} is
 shown on Fig. \ref{figtransloc} (a)  for various disorder strength $W$.
The corresponding slope $1/\xi_{loc}(W)$ is shown as a function of $W$ 
on Fig. \ref{figtransloc} (b) : we find that this slope 
vanishes linearly in $(W-W_c)$, in agreement with
 the exact result  \cite{Kun_Sou,MirlinBethe}
\begin{eqnarray}
\nu_{loc}=1
\label{nueq1}
\end{eqnarray}
and in agreement with Figure 7 of Ref. \cite{bell} 
concerning a random-scattering model on the Cayley tree.

\subsubsection{ Behavior of the typical transmission $T_{\infty}^{typ}$
in the delocalized phase $W<W_c$ near criticality }

\begin{figure}[htbp]
 \includegraphics[height=6cm]{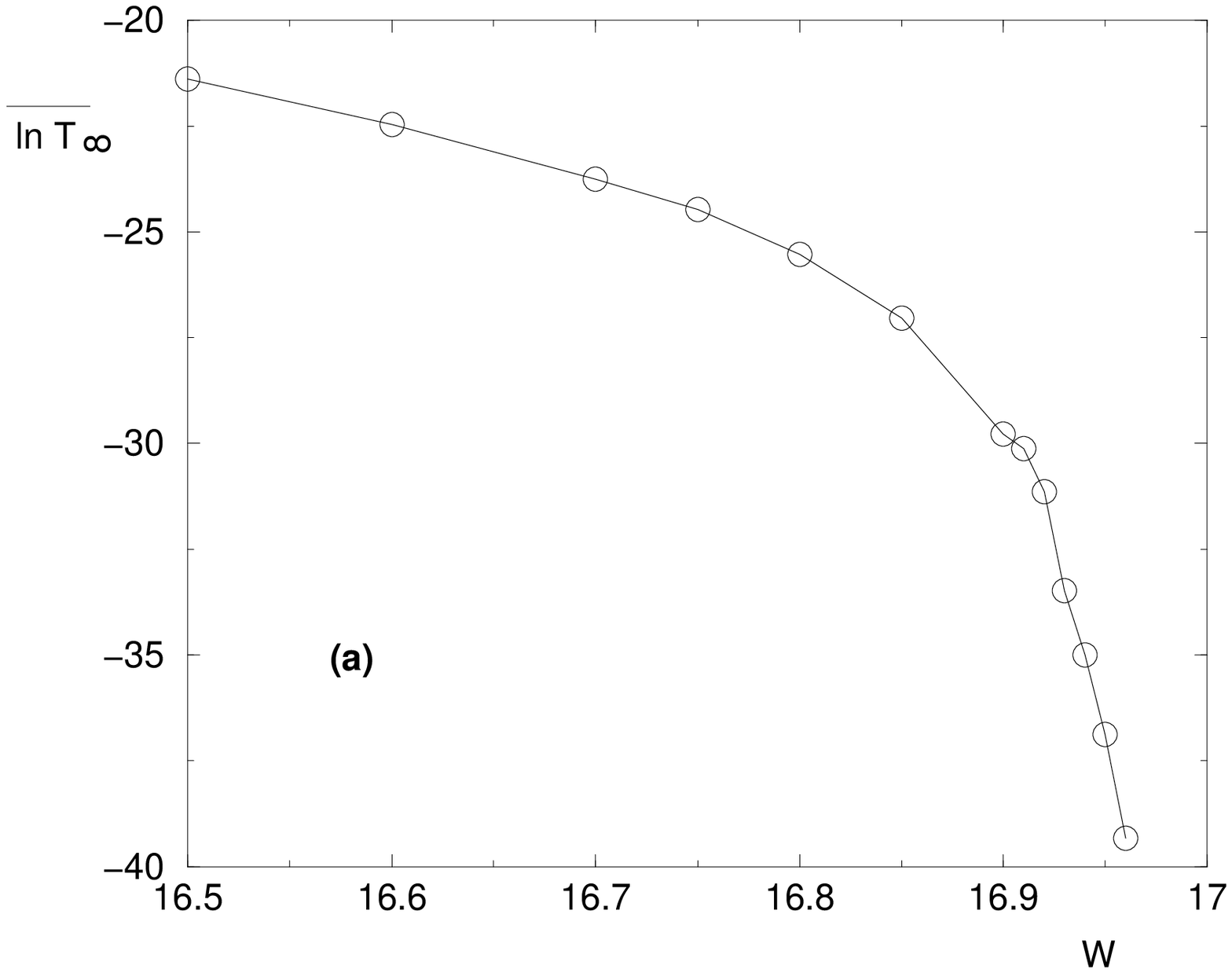}
\hspace{1cm}
\includegraphics[height=6cm]{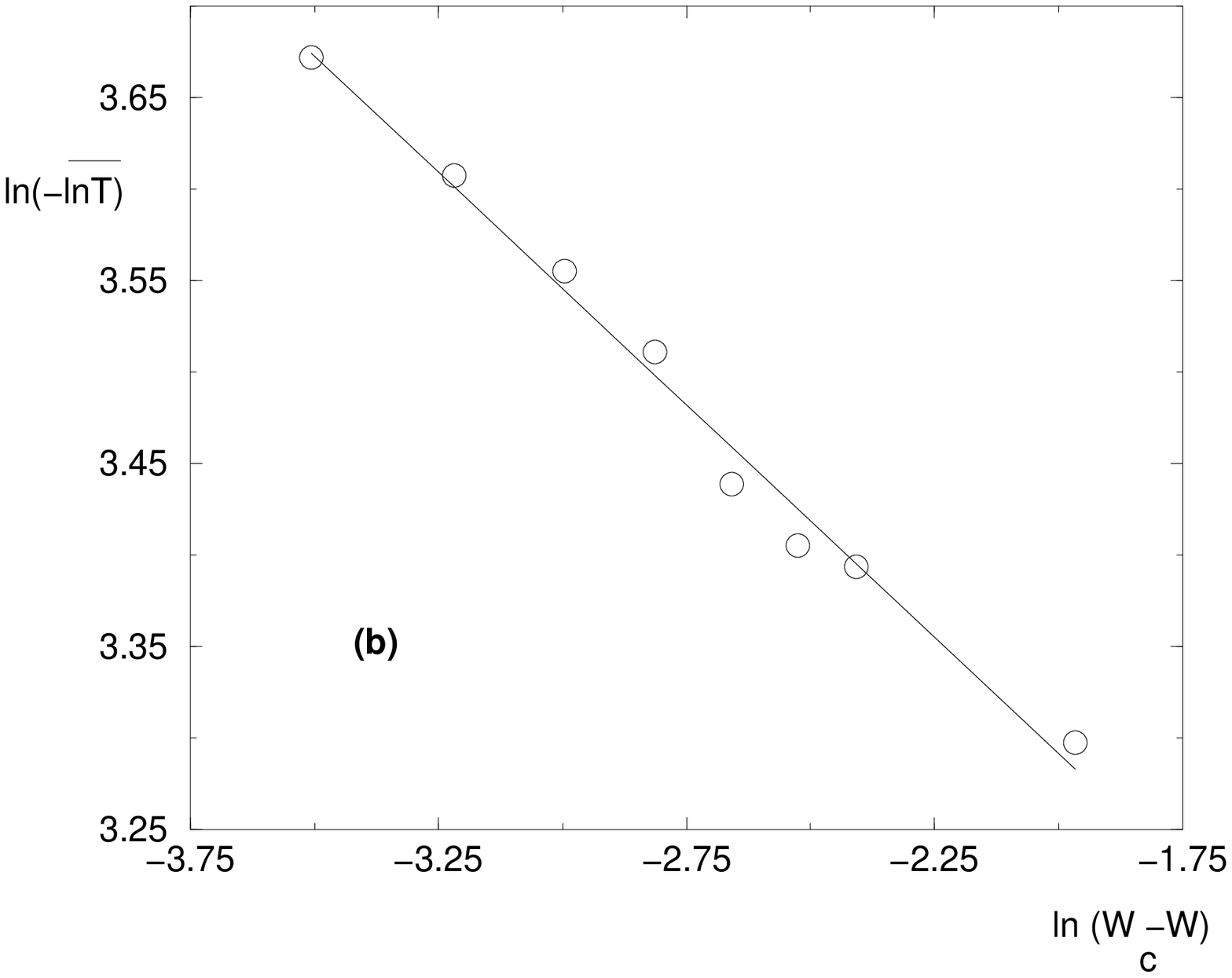}
\caption{ Behavior of the typical transmission $T_{\infty}^{typ}$
of the infinite tree in the delocalized phase :
(a)  $\ln T_{\infty}^{typ} \equiv \overline{ \ln T_{\infty}(W<W_c) }$ 
as a function of the disorder strength $W$
(b) same data in a log-log plot to measure the exponent of the essential singularity of Eq. \ref{essentialsingularitytransdeloc} :
 $ \ln (- \ln T_{\infty}^{typ} ) $ as a function of $\ln (W_c-W)$ :
the slope is of order $\kappa_T \sim 0.25 $
  }
\label{figtransdeloc}
\end{figure}

In the delocalized phase, the typical transmission remains finite 
in the limit where the number of generations $N$ diverges  
\begin{eqnarray}
\overline{ \ln T_N(W<W_c,N) } \opsimeq_{N \to \infty} 
 \overline{ \ln T_{\infty}(W<W_c) } > -\infty
\label{transdeloc}
\end{eqnarray}
As shown on Fig. \ref{figtransdeloc}, we measure the following 
 essential singularity behavior of the typical transmission
\begin{eqnarray}
\overline{ \ln T_{\infty}(W<W_c) }
 \opsimeq_{W \to W_c^-} -  (W_c-W)^{- \kappa_T}  \ \ \ 
 { \rm with } \ \  \kappa_T \sim 0.25
\label{essentialsingularitytransdeloc}
\end{eqnarray}

The presence of essential singularities in transport properties
near the localization transition on the Bethe lattice 
has been found in \cite{MirlinBethe} via the supersymmetric formalism
(see also \cite{zirnbauer,verbaar,efetov2lengths,efetovbook}
 where similar results have been obtained for the case where
 Anderson tight-binding model is replaced by a non-linear $\sigma$-model).
In particular, Eq. 71 of Ref. \cite{MirlinBethe})
states that the leading critical behavior of the diffusion constant
is given by : $ \ln {\cal D} \sim - \vert E-E_c \vert^{-1/2}$
 (Eq. 71 of Ref. \cite{MirlinBethe}). We note that the exponent
in this essential singularity is $1/2$ instead of the exponent
of order $1/4$ that we measure. The reason for this difference
could be that the results of Ref. \cite{MirlinBethe}
are based on the computation of the 
{ \it disorder-averaged two-point density-density correlation function }
(see Eq. 2 of Ref. \cite{MirlinBethe}), whereas our numerical 
results concern the { \it typical value of Landauer transmission},
and not the disorder-averaged transmission which is expected to be governed
by rare events (see below the section \ref{distritrans}).
Also it is not clear to us what exactly represents 
the diffusion constant computed in \cite{MirlinBethe},
because it is known that on the Bethe lattice
the dynamics is not diffusive but ballistic in the delocalized phase
\cite{klein}, and that more generally 
the random walk on the Bethe lattice is not diffusive
because the tree-geometry induces an
 effective bias away from the origin (see for instance
\cite{RWBethe} and references therein). As a consequence,
some implicit reinterpretation of the Bethe lattice  
seems to underlie the statements of Ref. \cite{MirlinBethe}
that makes difficult a precise comparison with our present numerical results.

\subsubsection{ Finite-size scaling in the critical region  }

\label{fsstrans}

If there exists some finite-size scaling in the critical region for
the typical Landauer transmission of the form
\begin{eqnarray}
\overline{ \ln T_N(W)} \opsimeq - N^{\rho_T} G \left( N^{1/\nu_T^{FS}} 
(W_c-W)  \right)
\label{transfss}
\end{eqnarray}
the matching of our results in the 
localized phase  (see Eq. \ref{transloc})
 and in the delocalized phase 
(Eq. \ref{essentialsingularitytransdeloc})
requires a finite-size correlation length exponent $\nu^{FS}_T$ of order
\begin{eqnarray}
\nu^{FS}_T = \nu_{loc}+\kappa_T=1+\kappa_T \simeq 1.25
\label{nuFStrans}
\end{eqnarray}
In another traveling/non-traveling phase transition studied in Ref \cite{simon}
(see the summary in Appendix C), it has been obtained
 that the finite-size scaling exponent 
$\nu_{FS}$ is determined by the relaxation rate towards the finite value
in the non-traveling phase. We have thus studied the relaxation length 
towards the finite value in the delocalized phase. 
We find that our data for
$\ln T_N$ are compatible with the form
\begin{eqnarray}
\overline{\ln T_{N} (W<W_c) } \simeq -  (W_c-W)^{- \kappa_T} a_N
\label{stationarry}
\end{eqnarray}
where $a_N$ is a random stationary process as a function of $N$.
We find that its autocorrelation function is exponential
\begin{eqnarray}
C(N)  \simeq e^{- \frac{ N }{\xi_{relax}(W)}}
\label{defcorre}
\end{eqnarray}
and we measure that the relaxation 
length $\xi_{relax}(W)$ diverges with an exponent
\begin{eqnarray}
 \xi_{relax}(W)\oppropto \frac{1}{(W_c-W)^{\nu_{relax}}} \ \ {\rm with }
\ \ \nu_{relax} \simeq 1.21
\label{taurelaxtrans}
\end{eqnarray}
of the order of the exponent $\nu^{FS}_T$ of Eq. \ref{nuFStrans}.
We thus obtain that the critical properties are qualitatively similar
to the critical properties described in Ref \cite{simon}
 (see the summary in Appendix C) : the traveling phase is characterized by
a velocity that vanishes linearly, but the finite size scaling is governed by
the relaxation length towards the asymptotic
 finite value of the non-traveling phase.
Exactly at criticality, we thus expects the following
stretched exponential decay of the typical transmission
\begin{eqnarray}
\overline{\ln T_{N} (W_c) } \simeq -  N^{\rho_T}
\label{transcriti}
\end{eqnarray}
where the exponent $\rho_T$ is related to the other exponents by
(see the scaling relations of Eqs \ref{kapparelation} and \ref{nutyprelation} in Appendix C)
\begin{eqnarray}
\rho_T = \frac{ \kappa_T}{\nu^{FS}_T} = 1- \frac{1}{\nu^{FS}_T}
\label{transrho}
\end{eqnarray}
From our previous estimate of the exponent $\kappa_T \simeq 0.25$, this would correspond to
the numerical value
\begin{eqnarray}
\rho_T = \frac{ \kappa_T}{1+\kappa_T} \simeq 0.2
\label{transrhonume}
\end{eqnarray}
We have not been able to measure this stretched exponential behavior exactly at criticality from our data,
because a precise measure of the exponent $\rho_T$ would require to be exactly at the critical point.

\subsubsection{ Distribution of the logarithm of the transmission  }

\label{distritrans}

\begin{figure}[htbp]
 \includegraphics[height=6cm]{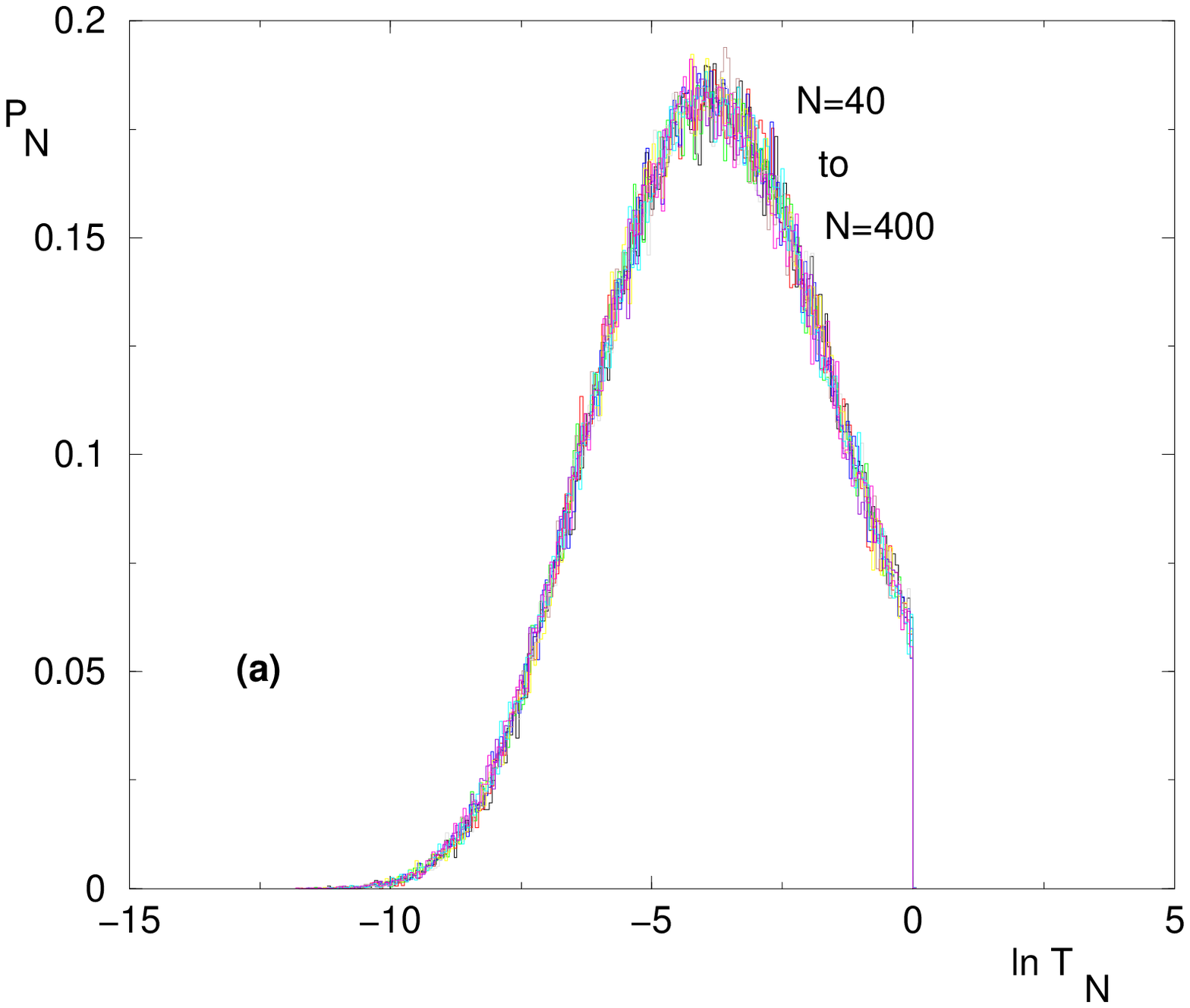}
\hspace{1cm}
 \includegraphics[height=6cm]{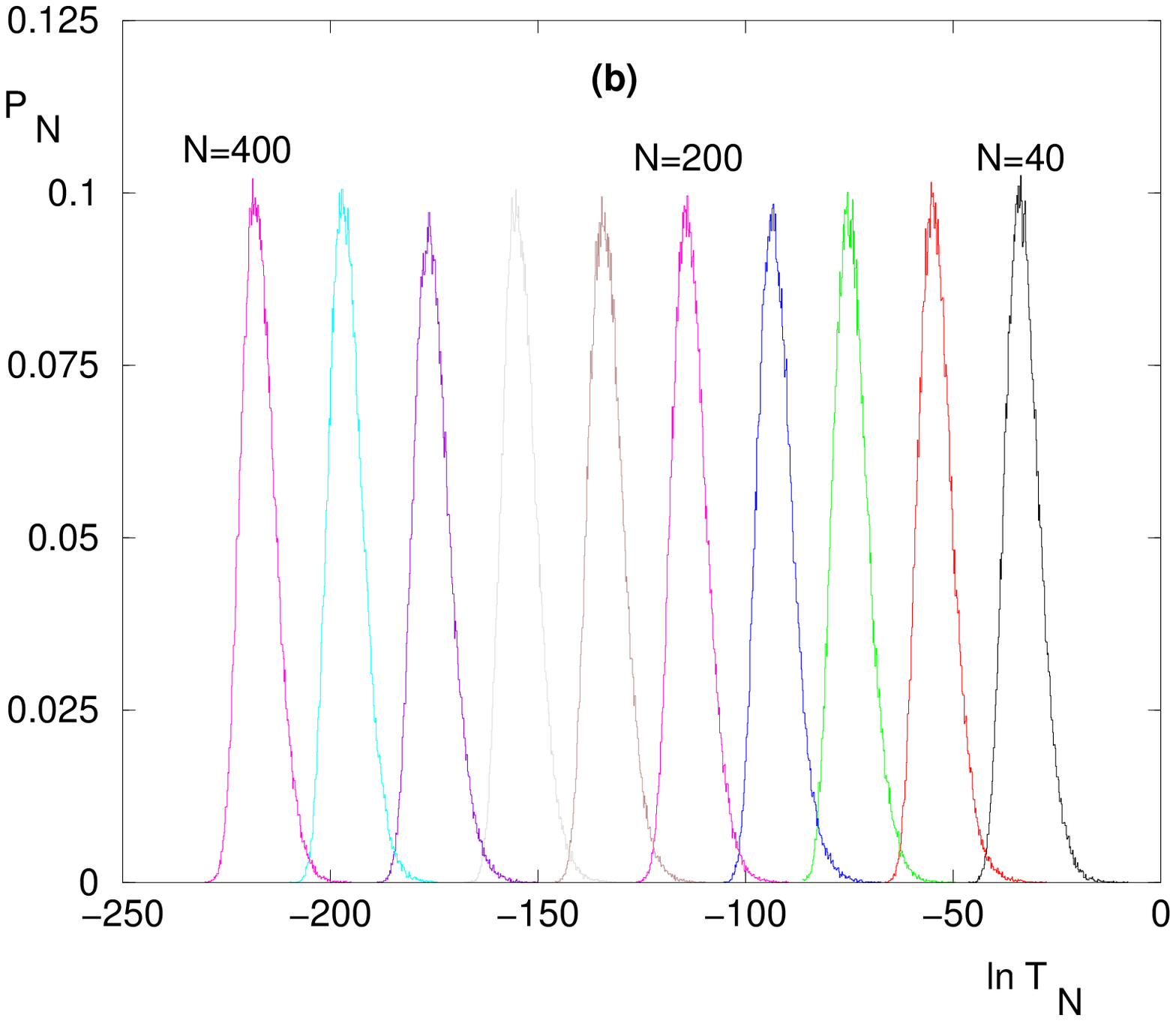}
\caption{ Evolution with $N$ of the probability distribution $P_N(\ln T_N)$
of the logarithm of Landauer transmission  :
(a)  In the delocalized phase (here $W=10$), the 
probability distribution $P_N(\ln T_N)$  does not move with $N$, and vanishes
with a discontinuity at the boundary $\ln T_N=0$.
(b) In the localized phase (here $W=25$), the 
probability distribution $P_N(\ln T_N)$ moves with $N$ as a traveling wave
of fixed shape.
  }
\label{fighistotrans}
\end{figure}

Up to now, we have only discussed the 
behavior of the typical transmission of Eq. \ref{defTtyp}
as a function of $N$ and $W$. We now turn to the probability distribution
of the logarithm of the transmission around its averaged value, 
i.e. we consider 
the distribution of the relative variable 
\begin{eqnarray}
u \equiv \ln T_N - \overline{ \ln T_N }
\label{variableu}
\end{eqnarray}
We find that as $N \to \infty$, 
this variable remains finite not only in the delocalized
phase where $\overline{ \ln T_{\infty} }$ is finite, 
but also in the localized phase
where $\overline{ \ln T_{N} }$ decays linearly in $N$ (Eq. \ref{transloc}).
This means that in the localized phase,
the probability distributions $P_N(\ln T_N )$ actually propagates
as a traveling wave with a fixed shape around its moving center
 $\overline{ \ln T_{N} }$
as shown on Fig. \ref{fighistotrans} (b).
This phenomenon has already been seen for random scattering models
on the Bethe lattice (see Fig. 8b of Ref. \cite{bell}).
This is in contrast to the broadening with $L$ observed in low dimensions
$d=1,2,3$ (see Eq. \ref{transd}) : the Cayley tree thus corresponds to 
 $\omega_{Cayley}=0$ in Eq. \ref{transd}.

\begin{figure}[htbp]
 \includegraphics[height=6cm]{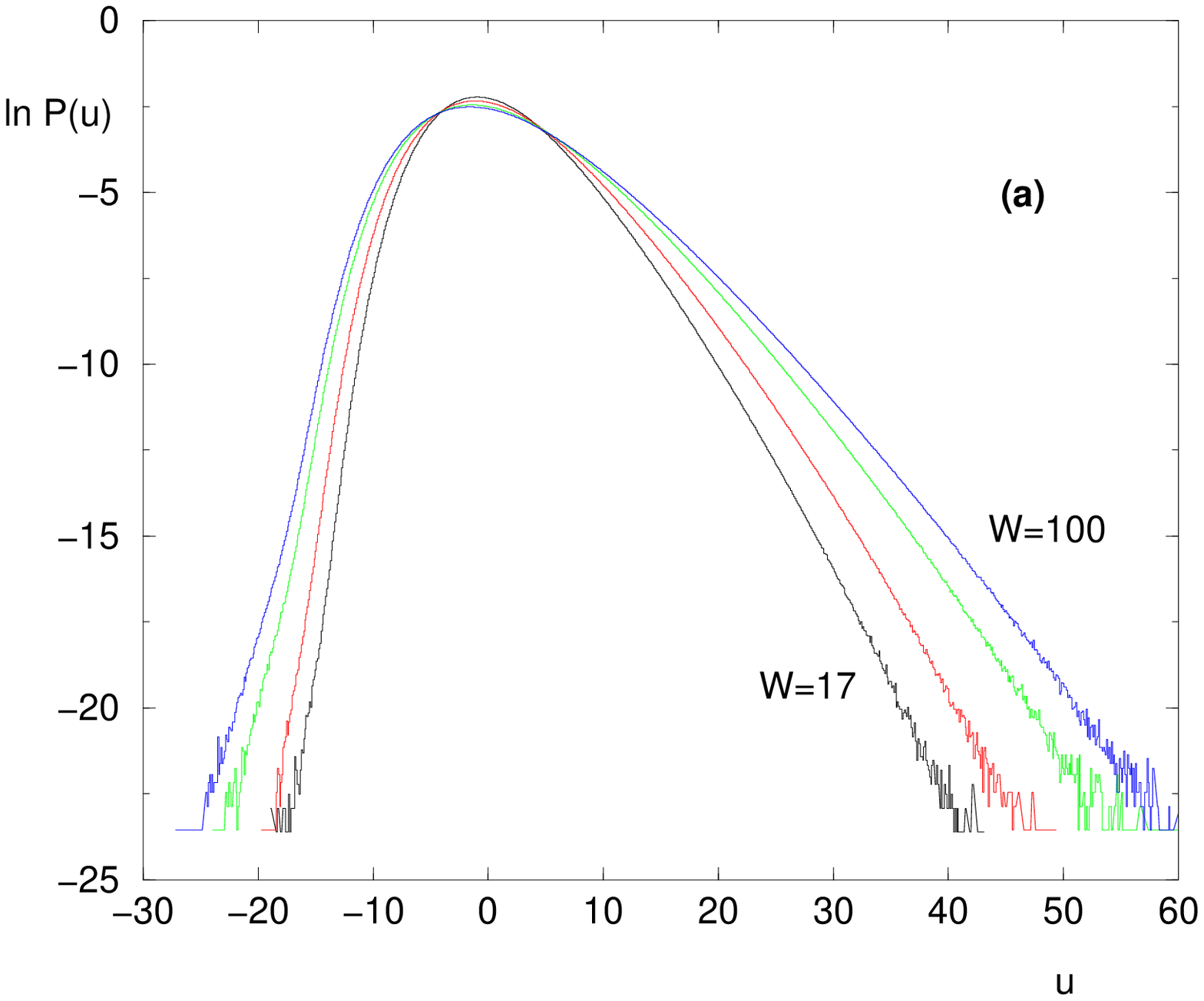}
\hspace{1cm}
 \includegraphics[height=6cm]{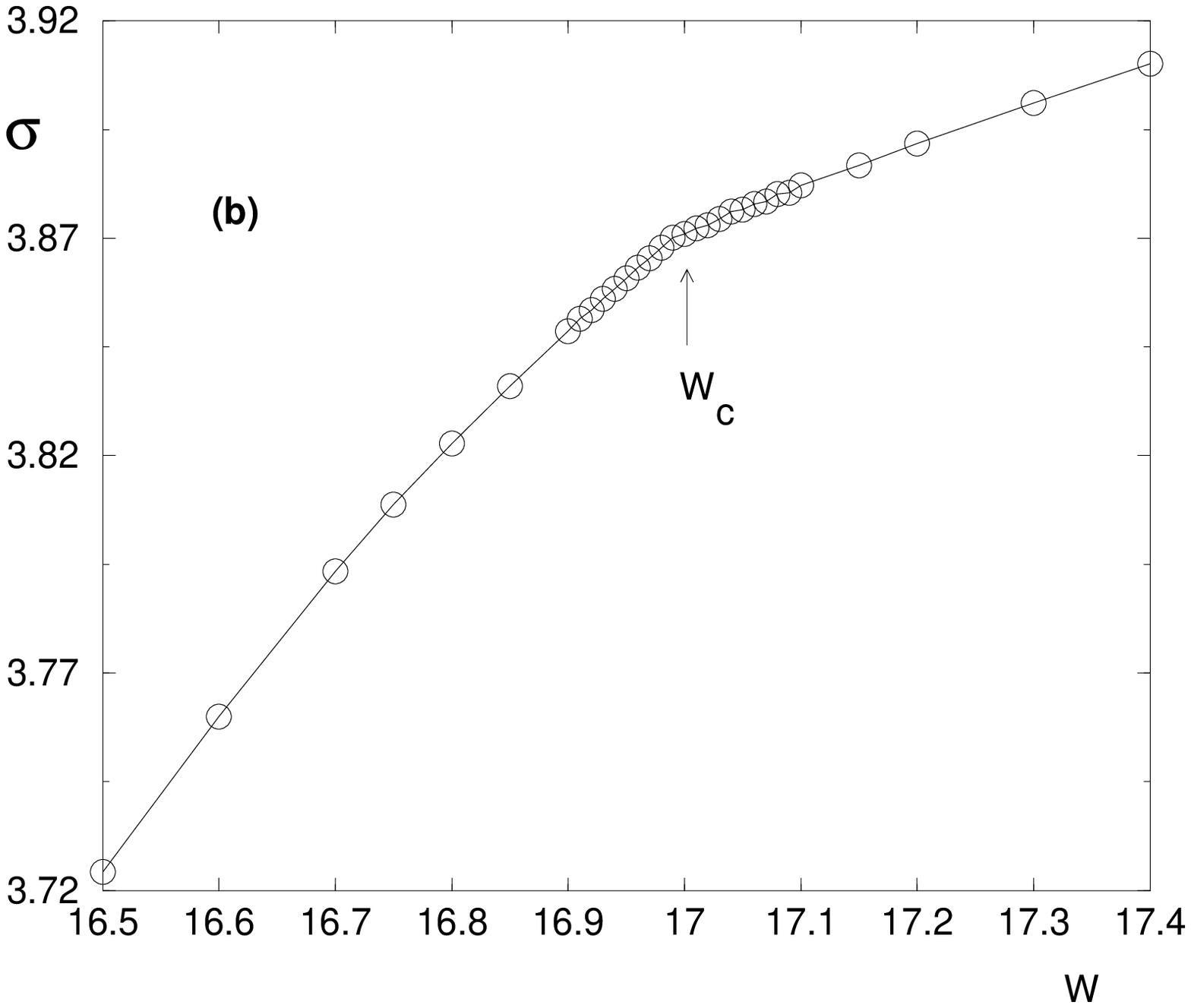}
\caption{ 
(a) Logarithm of the probability distribution $P(u)$ of 
$u = \ln T_N - \overline{ \ln T_N }$ as $W$ varies in the localized phase 
for various disorder strengths $W=17,25,50,100$ : 
the exponential decay as $e^{- \beta u}$ corresponds
to slopes of order $\beta(W=17) \simeq 0.48$, $\beta(W=25) \simeq 0.43$,
$\beta(W=50) \simeq 0.35$, $\beta(W=100) \simeq 0.33$.
(b) Width $\sigma= ( \overline{u^2})^{1/2}$ of the probability distribution of 
$u=\ln T_N - \overline{ \ln T_N }$ as a function of $W$ :
it remains finite both in the localized phase and in the delocalized phase,
but it presents a cusp singularity at criticality.
  }
\label{figtranswidth}
\end{figure}

The fact that the shape is fixed can be used numerically to measure
more precisely the tails of this probability distribution 
by accumulating data over iterations : we show on Fig. \ref{figtranswidth} (a)
the histograms of $u$ obtained by this procedure for various disorder strength
 $W$.
An essential property of this distribution is the right exponential tail
\begin{eqnarray}
P_W(u) \opsimeq_{u \to + \infty} e^{ -\beta(W) u }
\label{tailexpu}
\end{eqnarray}
In terms of the rescaled transmission $t=T_N/T_N^{typ} = e^{u}$ (see Eq. \ref{variableu}), this corresponds to the power-law decay
\begin{eqnarray}
P( t =T_N/T_N^{typ}  ) \opsimeq_{t \to + \infty} \frac{1}{ t^{ 1+\beta(W) } }
\label{tailpowert}
\end{eqnarray}
We find that the selected exponent 
$\beta_{selec}(W)$ slightly grows as the disorder strength $W$ decreases,
from a value of order $\beta(W=100) \simeq 0.33$ 
for the strong disorder $W=100$
towards a value of order
 $\beta(W=17) \simeq 0.48$ near criticality (see Figure \ref{figtranswidth} a).
The velocity of the traveling wave propagation
of the whole distribution is actually determined by this power-law tail,
as explained in detail in Appendix A.
In particular, the critical value $\beta(W \to W_c)=1/2$
 has been predicted in \cite{abou}
(see Eq. \ref{betacdemi} in  Appendix A).
An important consequence of the power-law tail of Eq. \ref{tailpowert}
with $\beta(W) \leq 1/2$ is that all integer moments $\overline{T_N^n}$
of the Landauer transmission will be governed by rare events.

Finally, we show on Fig. \ref{figtranswidth} (b) the
width of the probability distribution of $\ln T_N $ as a function of $W$ :
it remains finite both in the localized phase and in the delocalized phase,
but it presents a cusp singularity at criticality
because in the delocalized phase, the distribution presents
a singularity at finite distance from the typical value,
as is clearly visible on Fig. \ref{fighistotrans} (a) :
the transmission is bounded by $T \leq 1$, i.e. the histogram of
  $\ln T_N $ presents a discontinuity at $\ln T_N=0$.

\section{ Statistics of  eigenstates }

\label{eigen}

\begin{figure}[htbp]
 \includegraphics[height=6cm]{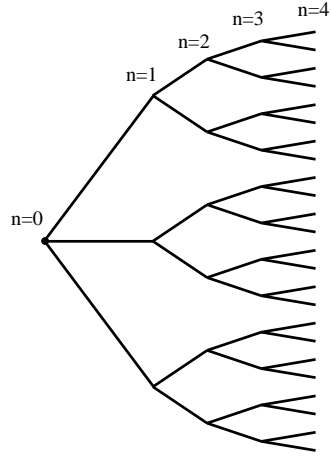}
\caption{ Cayley tree with branching ratio $K=2$ where each interior site
has $K+1=3$ neighbors. On the figure the tree ends at generation $2N=4$. }
\label{figtree}
\end{figure}

The Landauer transmission studied in the previous section is of course
the most appropriate observable to characterize the transport properties
and to find the transition between the conducting/non-conducting phases.
However, one expects that these transmission properties that emerge 
when the disordered sample is linked to incoming and outgoing wires 
are related to the nature of eigenstates of the disordered sample
in the absence of these external wires (see Fig. \ref{figtree}) 
To determine whether a normalized eigenstate 
\begin{eqnarray}
\sum_x \vert \psi(x) \vert^{2}=1
\label{normapsi}
\end{eqnarray}
is localized or delocalized, the usual parameters are 
 the Inverse Participation Ratios (I.P.R.)
\begin{eqnarray}
I_q \equiv \sum_x \vert \psi(x) \vert^{2 q}
\label{defIq}
\end{eqnarray}
of arbitrary power $q$. 
Another important quantity to characterize the spatial extent
of an eigenstate is its entropy
\begin{eqnarray}
S \equiv - \sum_x \vert \psi(x) \vert^{2}  \ln \vert \psi(x) \vert^{2}
\label{defS}
\end{eqnarray}
The normalization condition yields the idendity $I_{q=1}=1$, and the entropy 
corresponds to  
\begin{eqnarray}
S =  -  \partial_q I_q \vert_{q=1} = - \partial_q \ln I_q \vert_{q=1}
\label{SderiIq}
\end{eqnarray}
On an hypercubic lattice of size $L$ in dimension $d$, containing ${\cal N}_L=L^d$ sites,
the delocalized phase is characterized by the following typical behaviors 
\begin{eqnarray}
\overline{ \ln I_q } \sim - d (q-1) \ln L = - (q-1) \ln {\cal N}_L
\label{delocIqdimd}
\end{eqnarray}
and
\begin{eqnarray}
\overline{ S } \sim  d  \ln L =  \ln {\cal N}_L
\label{delocSdimd}
\end{eqnarray}
These scalings are the same for pure homogeneous eigenfunctions with
 $\vert \psi(x) \vert=1/L^d$ for all sites $x \in L^d$.
At criticality, the scalings of the I.P.R. involve a whole series of non-trivial exponents
and the wave function is said to be multifractal (see the review
 \cite{mirlinrevue}).
Finally in the localized phase, the I.P.R. and the entropy are finite.

In the present section, we discuss the statistical properties of these I.P.R.
and entropy for zero-energy eigenstates for the Cayley tree of Fig. \ref{figtree}
 as a function of the number $N$ of tree generations
and of the disorder strength $W$.

\subsection{  Reminder on the Miller-Derrida framework 
to construct eigenstates }

We refer to \cite{MD} where it is explained how eigenstates
of finite trees can be constructed and how the density of states
can be obtained. Note that on a tree, boundary sites dominate
so that one needs a substraction procedure to obtain the appropriate
bulk density of states.
In the following, we are not interested into the density of states,
which does not contain any information on the localized/delocalized nature
of the spectrum. We wish to study
 instead the spatial properties of eigenstates
of zero-energy $E=0$ (center of the band).
The Schr\"odinger equation \ref{schodingerinside} 
yields as before the recursion of 
Eq. \ref{riccatiinside} for the Riccati variables defined in Eq. \ref{riccati}.
The difference with the scattering case is now in the boundary conditions
\begin{eqnarray}
R(2N,j) = - \epsilon(2N,j) 
\label{riccatifirsteigen}
\end{eqnarray}
that replace Eq. \ref{riccatifirst}. As a consequence, 
the Riccati variables are now real
(and not complex).
The energy $E=0$ will indeed be an eigenstate only if
the Schr\"odinger equation is also satisfied at the center of the finite tree
 at the center site
\begin{eqnarray}
0=  \epsilon(0) + \sum_{m=1}^{K+1} \frac{1}{ R(des_m(0)) }
\label{riccaticenter}
\end{eqnarray}
Since the on-site energy $\epsilon(0)$ is a random variable
 drawn with some distribution,
we may consider that we choose $\epsilon(0)$ to satisfy Eq. \ref{riccaticenter}
to obtain a typical eigenstate of zero energy
 and to study its spatial properties.
 Let us first describe zero-energy 
eigenstate in the pure case to stress the peculiarities
of the tree geometry.

\subsection{ Structure of zero-energy eigenstate on the pure tree }

 In the pure case where all on-site energies vanish $\epsilon(n,j)=0$,
the zero-energy eigenstate with radial symmetry reads
\begin{eqnarray}
\psi_{2n-1} && =0 \\
\psi_{2n} && = \psi_0 \left( - \frac{1}{K} \right)^n
\label{eigenpur}
\end{eqnarray}
where the root amplitude $\psi_0$ is determined by the normalization condition
\begin{eqnarray}
1 && =  \sum_x \vert \psi(x) \vert^2
= \vert \psi_0 \vert^2 + (K+1) K \vert \psi_2 \vert^2 
+ (K+1) K^3 \vert \psi_4 \vert^2 + ...
 + (K+1) K^{2N-1} \vert \psi_{2N} \vert^2 \nonumber \\
&& = \vert \psi_0 \vert^2 \left[ 1+ \frac{K+1}{K} N \right]
\label{normapur}
\end{eqnarray}
This state is the analog of the scattering state described in section
 \ref{puretrans}.
In particular, one sees again an exponential decay of 
the wavefunction in this delocalized case
which is very peculiar to the tree geometry :
in the normalization condition of Eq. \ref{normapur}, this exponential decay
is exactly compensated by the exponential growth of the number of sites
at generation $2n$, so that all generations $2n=2,4,..2N$ carry exactly
the same weight in the normalization condition, and the root 
weight $\vert \psi_0 \vert^2$ vanishes as $1/N$ in the thermodynamic limit,
where $(2N)$ is the number of generations of the tree.

For this state, the usual Inverse Participation Ratio of Eq. \ref{defIq}
 for $q=2$ reads
\begin{eqnarray}
I_2^{pur}(2N) && =  \sum_x \vert \psi(x) \vert^4
= \vert \psi_0 \vert^4 + (K+1) K \vert \psi_2 \vert^4 
+ (K+1) K^3 \vert \psi_4 \vert^4 + ... + (K+1) K^{2N-1} \vert \psi_{2N} \vert^4 \\
&& = \vert \psi_0 \vert^4 \left[ 1+ \frac{K+1}{K^3} \ \frac{1-\left(  \frac{1}{K^2} \right)^N}{1-  \frac{1}{K^2}} \right]
\label{ipppur}
\end{eqnarray}
More generally for $q >1$, one obtains the following decay with the number of generations
\begin{eqnarray}
I_q^{pur}(2N) \oppropto_{N \to \infty} \vert \psi_0 \vert^{2q} \sim \frac{1}{N^q} \ \ 
{ \rm  for } \ \  q>1
\label{ipppurq}
\end{eqnarray}
In terms of the total number of sites
\begin{eqnarray}
{\cal N}(2N) &&
 =   1+(K+1) \sum_{n=1}^{2N-1} K^n = 1+(K+1) \frac{K^{2N}-1}{K-1}
=  \frac{K^{2N} (K+1)-2}{K-1}
\label{calncaspur}
\end{eqnarray}
this corresponds to
\begin{eqnarray}
\ln I_q^{pur}(2N) \oppropto_{N \to \infty} -q \ln N \oppropto_{N \to \infty} -q \ln (\ln {\cal N}(2N) )  \ \ 
{ \rm  for } \ \  q>1
\label{ipppurcalN}
\end{eqnarray}
This behavior is thus very anomalous with respect to the corresponding decay
as $( - (q-1) \ln {\cal N}_L)$ for the I.P.P. of pure states 
in finite dimension $d$ (see Eq. \ref{delocIqdimd}).

Moreover, since the eigenfunction
 normalization yields the idendity $I_{q=1}=1$, 
one sees that there
exists some discontinuity as $q \to 1$. In particular, the entropy of Eq. \ref{defS}
which can usually be computed as the derivative of Eq. \ref{SderiIq}
should be computed directly here, and one obtains
\begin{eqnarray}
S^{pur}(2N) && = - \vert \psi(0)\vert^2 \ln \vert \psi(0)\vert^2 
- (K+1) K \vert \psi(2)\vert^2 \ln \vert \psi(2)\vert^2 -...
- (K+1) K^{2N-1} \vert \psi(2N)\vert^2 \ln \vert \psi(2N)\vert^2 \\
&& =  \vert \psi(0)\vert^2 
\left[ - \left(1+\frac{K+1}{K} N \right) \ln \vert \psi(0)\vert^2
+ \frac{K+1}{K} N (N+1) \ln K \right]
\label{entropiecaspurcalcul}
\end{eqnarray}
Using Eq \ref{normapur}, one obtains the following behavior for large $N$
\begin{eqnarray}
S^{pur}(2N) &&
 =     \ln \left( 1+\frac{K+1}{K} N\right)
+ \frac{\frac{K+1}{K} N (N+1) \ln K}{ \left( 1+\frac{K+1}{K} N\right)} 
 \opsimeq_{N \to \infty}  N \ln K
\label{entropiecaspur}
\end{eqnarray}
In terms of the total number of sites of Eq. \ref{calncaspur},
 the entropy is proportional to the logarithm of
 the number of sites
\begin{eqnarray}
S^{pur}(2N)  \opsimeq_{N \to \infty} 
 N \ln K \opsimeq_{N \to \infty} \frac{1}{2} \ln {\cal N}(2N) 
\label{entropiecaspurfin}
\end{eqnarray}
that should be compared with the growth as $\ln {\cal N}(2N) $ 
for pure eigenstates in dimension $d$ (see Eq. \ref{delocSdimd}).

In conclusion, the tree geometry induces very anomalous scalings 
in $(\ln \ln {\cal N})$ for the logarithm of I.P.R. of pure states
(see Eq. \ref{ipppurcalN}) with respect to the case of finite dimension $d$, 
whereas the entropy is well behaved in
 $(\ln {\cal N})$ (see Eq. \ref{entropiecaspurfin}).
In the disordered case, we expect to observe similar behaviors,
whereas the I.P.R. and the entropy will remain finite in the localized phase.

\subsection{ Recursion relation for Inverse Participation Ratios (I.P.R.)  }

To compute the Inverse Participation Ratio $I_q$ (Eq. \ref{defIq})
via recursion, one needs to introduce besides the Riccati variable of Eq. \ref{riccati}
the auxiliary variables defined by
\begin{eqnarray}
C^{(q)} (n,j) = \vert \frac{\psi(n,j)}{\psi(anc(n,j))} \vert^{2q}
+ \sum_{ l }
\vert \frac{\psi(l)}{\psi(anc(n,j))} \vert^{2q}
\label{defcq}
\end{eqnarray}
where the sum is over all sites $l$ that are descendants of the site $(n,j)$
of the tree.
These variables satisfy the following recurrence inside the tree
 $1 \leq n \leq 2N-1$
\begin{eqnarray}
C^{(q)} (n,j) = \left( \frac{1}{R^2(n,j)} \right)^q
\left[ 1+  \sum_{m=1}^{K} C^{(q)} (des_m(n,j)) \right]
\label{reccq}
\end{eqnarray}
and the initial conditions at the boundaries
\begin{eqnarray}
C^{(q)} (2N,j) = \left( \frac{1}{R^2(2N,j)} \right)^q
\label{cqinitiale}
\end{eqnarray}
At the central root, on needs to impose the normalization
\begin{eqnarray}
1= \sum_x \vert\psi(x) \vert^2
= \vert \psi(0) \vert^2 \left[ 1+ \sum_{m=1}^{K+1}
 C^{(1)} (1,m)\right]
\label{normapsi2}
\end{eqnarray}
that determines the weight $\vert \psi(0) \vert^2$ of the root in terms of the 
variables $C^{(1)}$ of the branches.

To compute the inverse participation ratio of parameter $q$ of Eq. \ref{defIq},
one needs the variables $C^{(q)}$ together with the variables $C^{(1)}$
\begin{eqnarray}
I_q= \sum_x \vert\psi(x) \vert^{2q}
= \vert \psi(0) \vert^{2q} \left[ 1+ \sum_{m=1}^{K+1}
 C^{(q)} (1,m)\right]
= \frac{\left[ 1+ \displaystyle \sum_{m=1}^{K+1}
 C^{(q)} (1,m)\right]}{\left[ 1+ \displaystyle \sum_{m=1}^{K+1}
 C^{(1)} (1,m)\right]^q }
\label{iqtree}
\end{eqnarray}

\subsection{ Recursion for the entropy  }

To compute recursively the entropy of Eq. \ref{defS}, one needs similarly to 
introduce the auxiliary variable 
\begin{eqnarray}
\sigma (n,j) = - \sum_{l}
\frac{ \vert \psi_l \vert^2}{ \sum_{l'} \vert \psi_{l'} \vert^2} \ln 
\frac{ \vert \psi_l \vert^2}{ \sum_{l'} \vert \psi_{l'} \vert^2} 
\label{defsigma}
\end{eqnarray}
where the sum over $l$ denotes the sum over the site $(n,j)$
and all its descendants : $\sigma (n,j)$ thus 
represents the entropy for the branch containing
$(n,j)$ and its descendants.

The initial conditions at the boundaries are simply
\begin{eqnarray}
\sigma (2N,j) = 0
\label{sigmainitiale}
\end{eqnarray}
and the recursion inside the tree $1 \leq n \leq 2N-1$ can be written as
\begin{eqnarray}
\sigma (n,j) = \frac{\sum_{m=1}^{K} C^{(1)} (des_m(n,j)) \sigma (des_m(n,j))}
{1+\sum_{m=1}^{K} C^{(1)} (des_m(n,j))} +S_{mix}
\label{pondere}
\end{eqnarray}
where the first term represents the weighted
 contribution of the branches entropies
and where the second term represents the mixing entropy
\begin{eqnarray}
S_{mix}  = -  p_0 \ln p_0 - \sum_{m'=1}^K p_{m'} \ln p_{m'}
\label{recsmix}
\end{eqnarray}
with the weights
\begin{eqnarray}
p_0= \frac{1}{(1+\sum_{m=1}^{K} C^{(1)} (des_m(n,j)))} \\
p_{m'} = \frac{C^{(1)} (des_{m'}(n,j)))}
{(1+\sum_{m=1}^{K} C^{(1)} (des_m(n,j)))} \\
\label{p0}
\end{eqnarray}
normalized to $p_0+\sum_{m'=1}^K p_{m'}=1$.
At the central site of the tree, one uses the same
 formula but with $(K+1)$ branches instead of $K$ branches.

\subsection{ Numerical results on the statistics of eigenstates }

\subsubsection{ Numerical pool method } 

To study the statistical properties of eigenstates,
we have used again the pool method explained in \ref{poolmethod}.
The only difference is that for the transmission, we have followed
the recursions for the complex Riccati variables, whereas here
 we follow
the recursions for the real Riccati variable $R$ and
 for the auxiliary variables
$C^{(1)}$, $C^{(2)}$ and $\sigma$ described above.

\subsubsection{ Statistics of the entropy of an eigenstate }

\begin{figure}[htbp]
 \includegraphics[height=6cm]{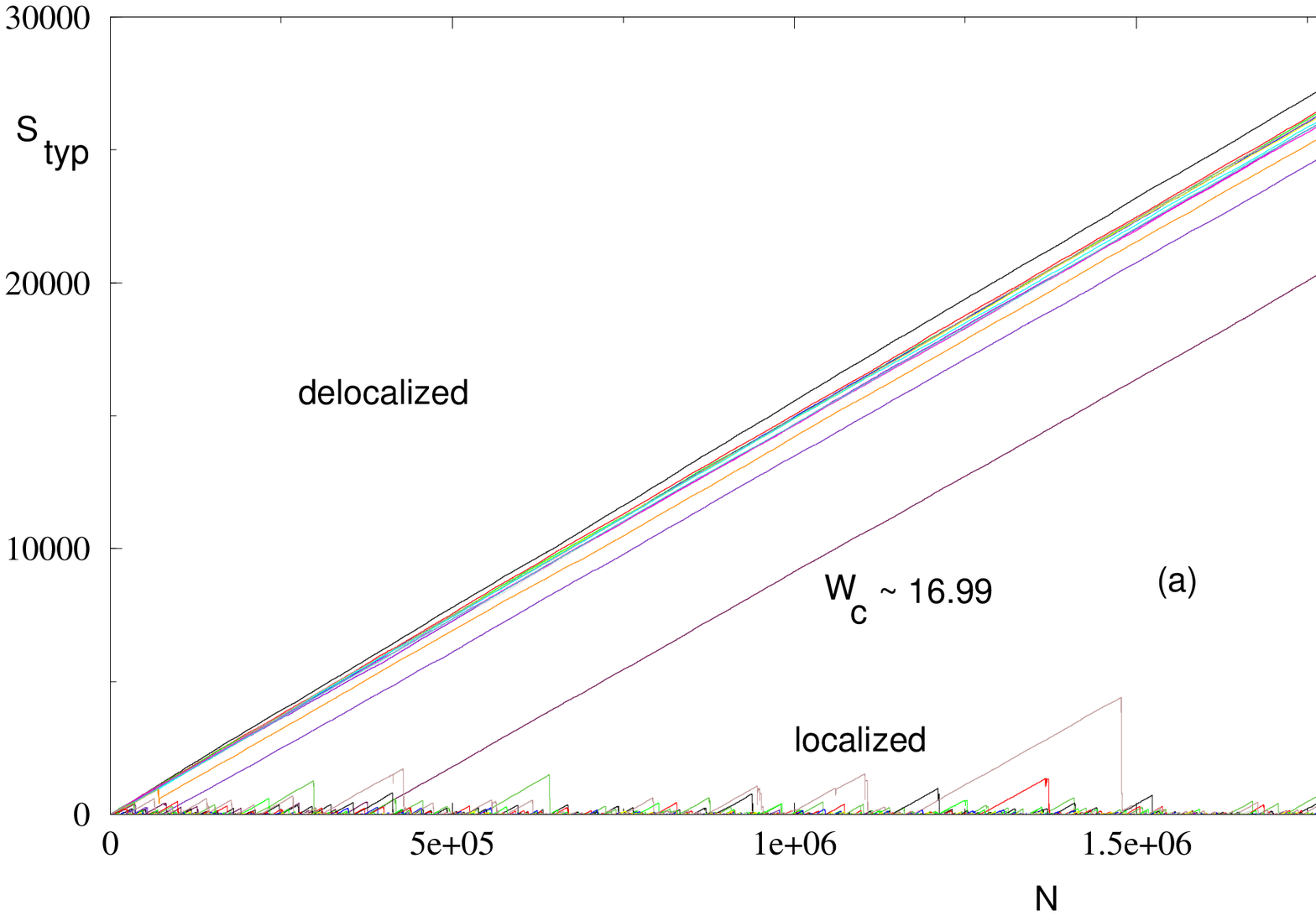}
\hspace{2cm}
\includegraphics[height=6.8cm]{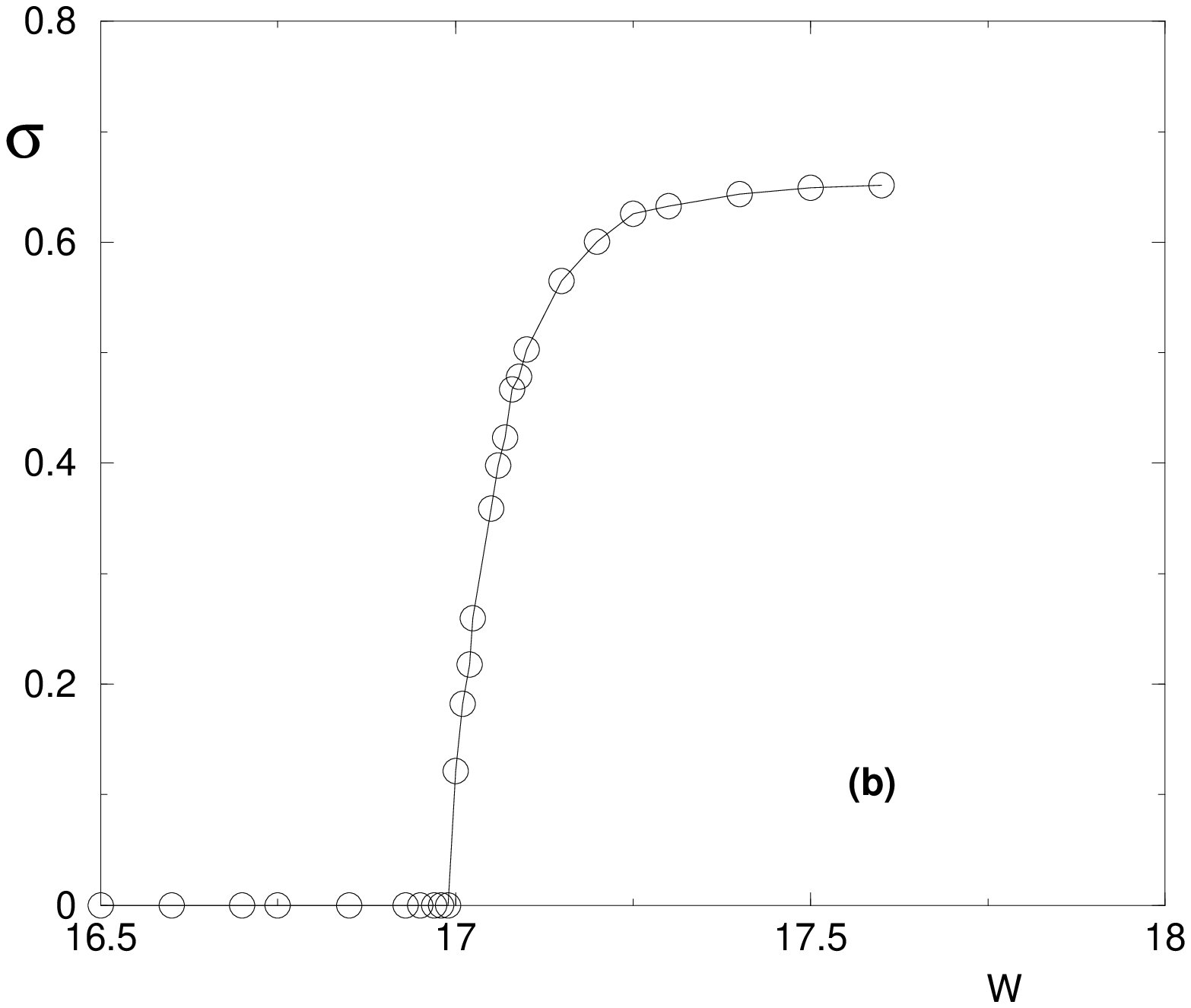}
\caption{ Statistics of the entropy 
$S(N) \equiv - \sum_x \vert \psi(x) \vert^{2}  \ln \vert \psi(x) \vert^{2} $
 of a normalized zero-energy eigenstate  :
(a) Evolution of the typical entropy 
$S_{typ}(N) \equiv e^{\overline{\ln S(N)}}$
with the number $N$ of generations :
it grows linearly in $N$ in the delocalized phase,
whereas it remains finite in the localized phase.
(b) Asymptotic width $\sigma= (\overline{v^2})^{1/2}$
of the relative variable $v= \ln S - \overline{\ln S}$
as a function of the disorder strength $W$ : 
the width converges to $0$ (as $1/N$) in the delocalized phase,
whereas it is finite in the localized phase}
\label{figEntropy}
\end{figure}

We first consider how the eigenstate entropy defined in Eq. \ref{defS} 
evolves with the number $N$ of generations.
As shown on Fig. \ref{figEntropy} (a), the typical value grows linearly in $N$
in the delocalized phase 
\begin{eqnarray}
S_{typ}(N,W<W_c) \equiv e^{\overline{\ln S(N)}} \opsimeq_{N \to \infty} a N
\label{entropydeloc}
\end{eqnarray}
where the factor $a$ varies smoothly 
with $W$ and does not vanish continuously near the critical point.
In this delocalized phase, we moreover find that the width
of the relative variable $v= \ln S_N - \overline{\ln S_N}$ 
decays to zero as $N \to \infty$ 
\begin{eqnarray}
\left[  \overline{ (\ln S_N - \overline{\ln S_N})^2 } \right]^{1/2}_{W<W_c}
  \opsimeq_{N \to \infty} \frac{1}{N} 
\label{entropywidthdeloc}
\end{eqnarray}
In the localized phase on the contrary, the typical value and 
the width of the relative variable $v= \ln S_N - \overline{\ln S_N}$
remain finite as $N \to +\infty$ (see Fig. \ref{figEntropy} (a) and (b)).

\begin{figure}[htbp]
 \includegraphics[height=6cm]{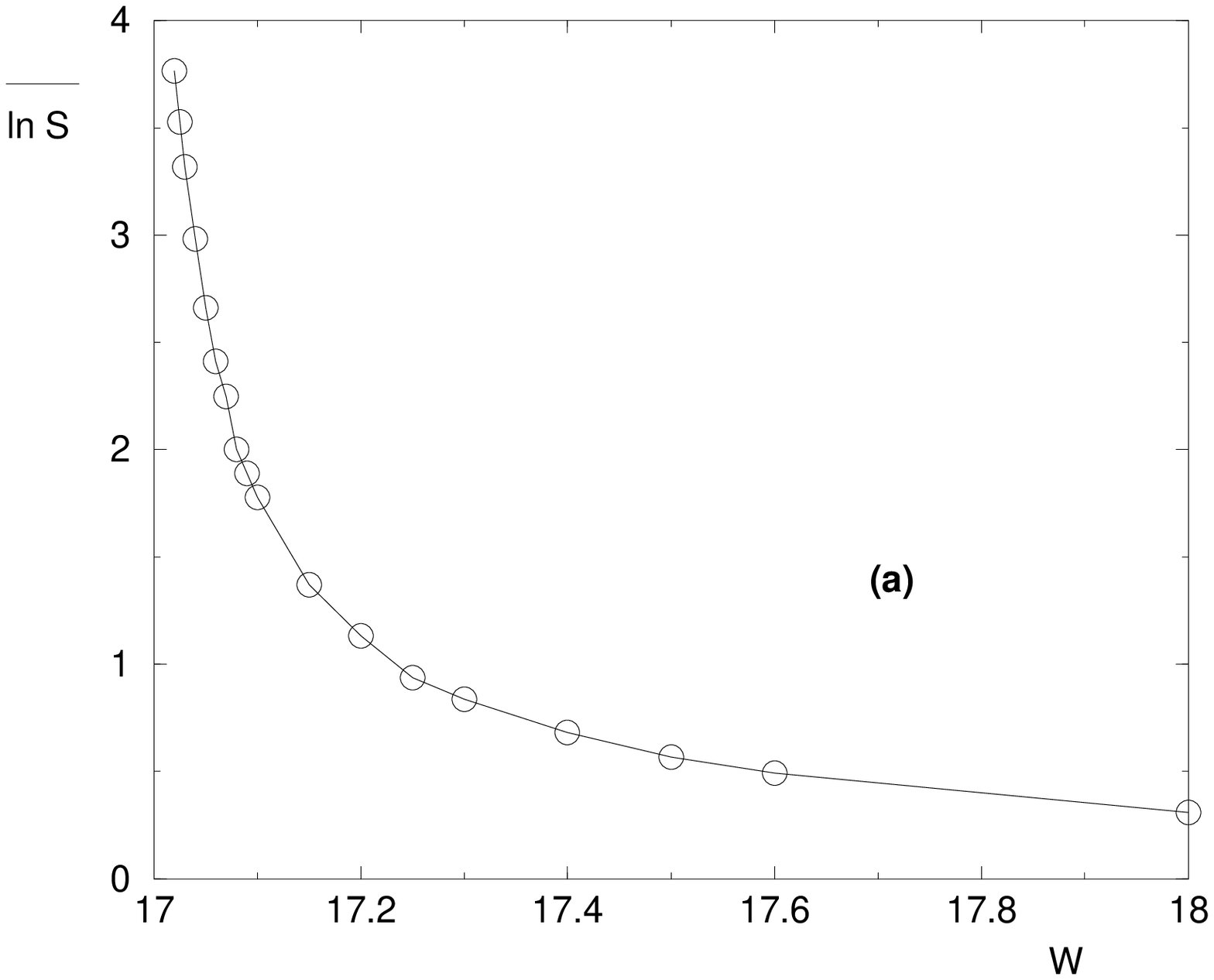}
\hspace{1cm}
\includegraphics[height=6cm]{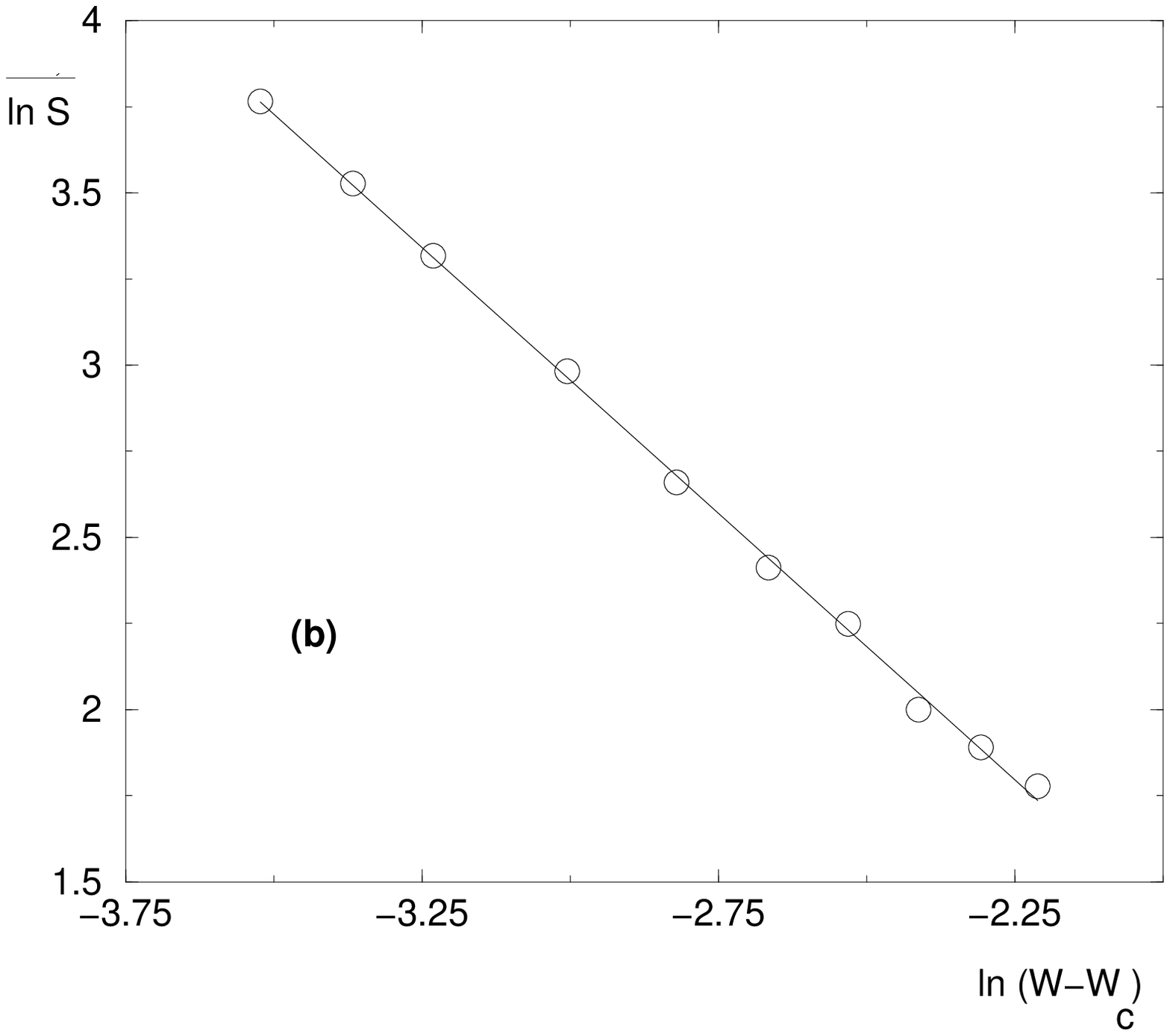}
\caption{ Critical behavior of the typical entropy 
$S_{typ}(N=\infty) = e^{\overline{ \ln S(N=\infty) }}$
 in the localized phase 
(a) $\ln S_{typ}(N=\infty) = \overline{ \ln S(N=\infty) }$ 
as a function of the disorder strength $W$ near the critical point.
(b) same data as a function of $\ln (W-W_c)$ 
to measure the exponent of the power law of Eq. \ref{critientropy64loc} :
the slope is of order $\nu_S \simeq 1.5 $.}
\label{fig64loc}
\end{figure}

As shown on Fig. \ref{fig64loc}, we find that the typical entropy
of an infinite tree of the localized phase diverges with the 
following power law near criticality
\begin{eqnarray}
 S_{typ} \equiv e^{\overline{ \ln S(N=+\infty) }}
 \opsimeq_{W \to W_c^+} \frac{1}{(W-W_c)^{\nu_S}} \ \ {\rm with}
\ \ \nu_S \simeq 1.5
\label{critientropy64loc}
\end{eqnarray}
The entropy measures the size of the region where the weight
$\vert \psi(x) \vert^2$ is concentrated, whereas 
the localization length $\xi_{loc} \sim 1/(W -W_c)^{\nu_{loc}=1}$
describes the far exponential
decay of the transmission $T$ (Eq. \ref{transloc}).
We thus conclude that
in the localized phase, besides the localization length
$\xi_{loc} \sim 1/(W -W_c)^{\nu_{loc}=1}$ known since \cite{Kun_Sou},
 there exists a larger diverging length scale
\begin{eqnarray}
\xi_S (W>W_c)  \oppropto_{W \to W_c^+} 
 \frac{1}{(W-W_c)^{\nu_s \sim 1.5}}
\label{xiSloc}
\end{eqnarray}
that characterizes the size where the weight $\vert \psi(x) \vert^2$
is concentrated.

\subsubsection{ Statistics of the 
root weight $ \vert \psi_N(0) \vert^2$ and of the I.P.R. $I_2$ }

As explained above, the root weight $ \vert \psi_N(0) \vert^2$ 
and of the I.P.R. $I_2$
are determined by the auxiliary variables $C^{(q)}$ introduced
 in Eq. \ref{defcq}.
We thus expect that the localized and delocalized phases
 correspond to the following behaviors
for the auxiliary variables $C^{(q)}$

(i) in the localized phase, the auxiliary variables $C^{(q)}$
 will remain finite random 
variables as the number of generations diverge $N \to \infty$.
 Then the root weight 
$\vert \psi_N(0) \vert^2$ and the I.P.R. $I_q$ remain finite as $N \to \infty$.

(ii) in the delocalized phase, the auxiliary variables $C^{(q)}$
 will instead grow
exponentially with the number $N$ of generations 
with some Lyapunov exponents $\lambda_q>0$ defined by
\begin{eqnarray}
 \overline{ \ln C^{(q)} } \opsimeq_{N \to \infty} \lambda_q N
\label{LyapunovCq}
\end{eqnarray}
The corresponding typical behaviors of the root weight and of the
 I.P.R. then reads
\begin{eqnarray}
 \overline{ \ln \vert \psi_N(0) \vert^2 } \opsimeq_{N \to \infty} - \lambda_1 N
\label{Lyapunovpsi0}
\end{eqnarray}
and
\begin{eqnarray}
 \overline{ \ln I_q } \opsimeq_{N \to \infty}  - ( q \lambda_1 - \lambda_q ) N
\label{LyapunovIq}
\end{eqnarray}

We find numerically that the probability distributions
of auxiliary variables 
$(\ln C^{q})$ move as traveling waves of velocity $\lambda_q$
 (see Eq. \ref{LyapunovCq}) with fixed shape 
(see Figure \ref{fighistoc1c2} of Appendix B 
where the fixed shape is shown for $q=1$ and $q=2$).
We explain in Appendix B how the Lyapunov exponents
 $\lambda_q$ can be determined
via a tail analysis that yields the identity (see Eq. \ref{selectionqidentity})
\begin{eqnarray}
\lambda_q =  q \lambda_1 
\label{LyapunovIqidentity}
\end{eqnarray}
so that the logarithm of the I.P.R. $I_q$ 
decays slower than linearly in $N$, because the coefficient
in Eq. \ref{LyapunovIq} exactly vanishes.
Physically, one could expect the decay to
be of order $(\ln N)$ as in Eq. \ref{ipppurcalN}
 concerning pure states on the Cayley tree,
but we have not been able to measure
 the behavior of the I.P.R., because within the pool method
that we use, it turns out that the 
pool-dependent critical point is not the same
for the variables $C^{(1)}$ and for $C^{(2)}$
 (see more details in section \ref{fspool}),
so that it does not seem easy with
 the pool method to extract reliable results concerning $I_2$.

In the remaining of this section, we thus focus on the statistical properties
 of the root weight $\vert \psi(0) \vert^2$
that is determined by the auxiliary variable $C^{(1)}$ via Eq. \ref{normapsi2}.
The fact that the pool-dependent critical point for $C^{(1)}$ is the same
as the critical point found for the Landauer transmission can be understood
from the tail analysis of the Appendices A and B that involve exactly the same integral kernel.

\begin{figure}[htbp]
 \includegraphics[height=5cm]{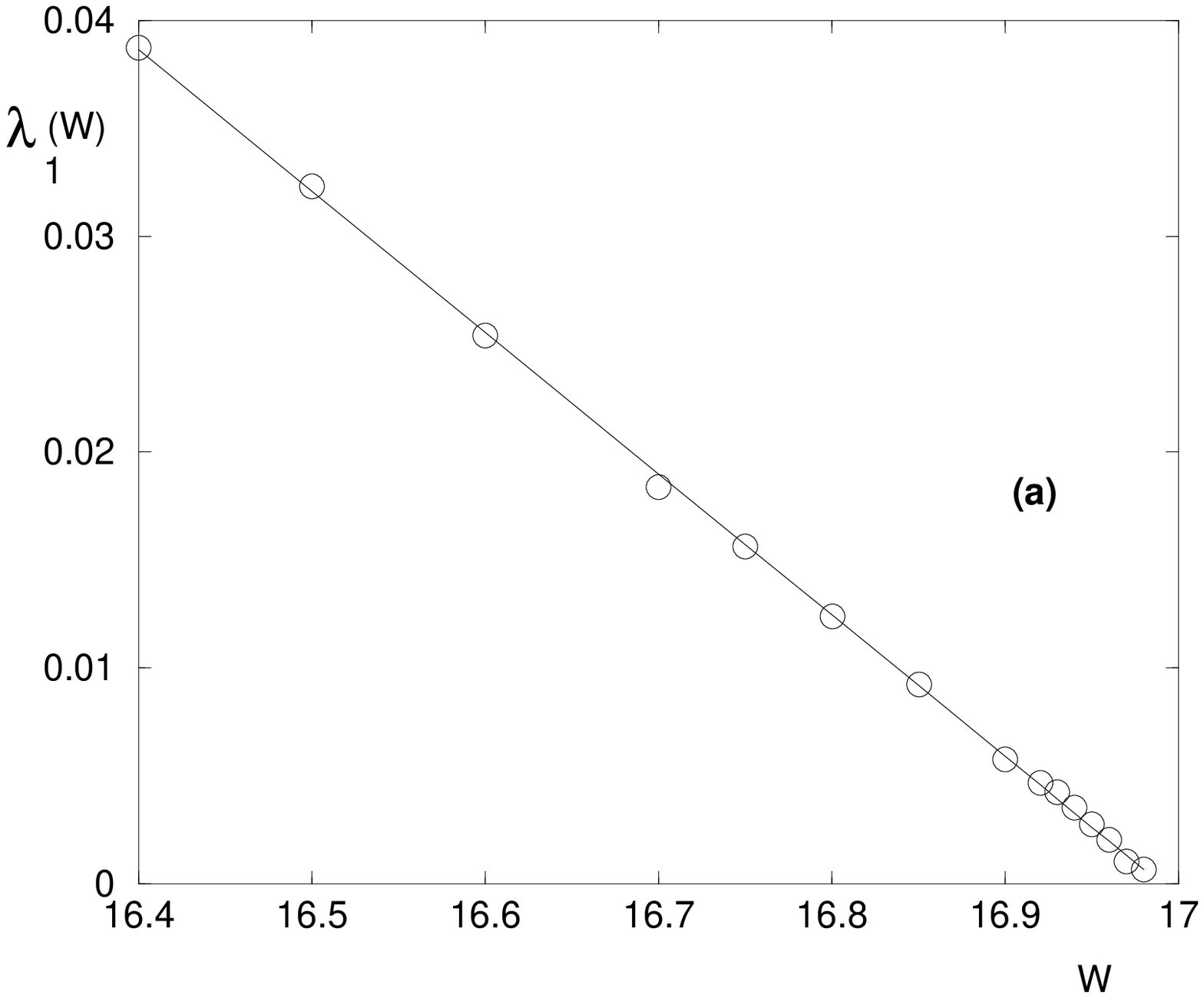}
\hspace{1cm}
 \includegraphics[height=5cm]{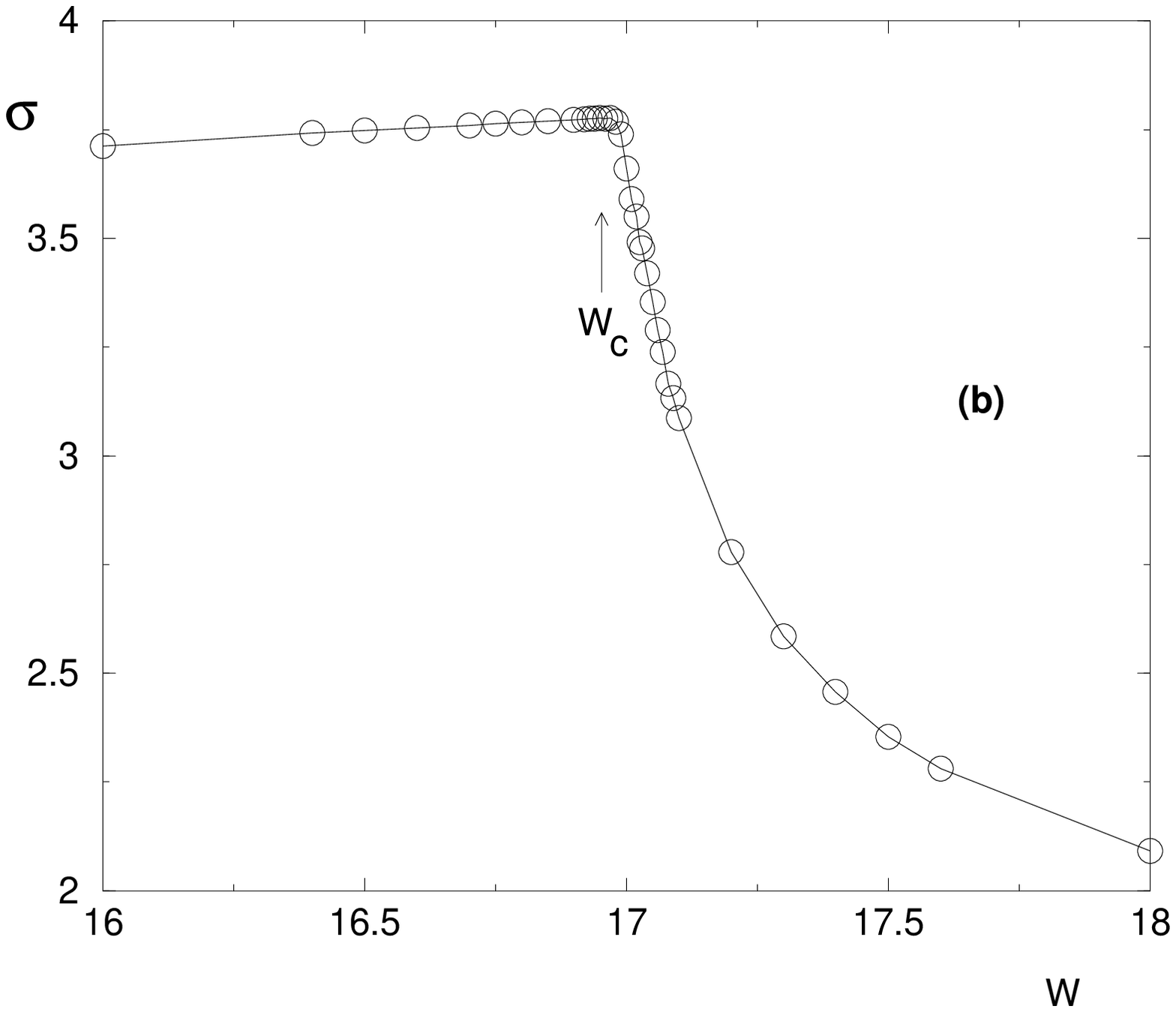}
\caption{ (a) In the delocalized phase,
the logarithm of the root weight decays linearly
$\overline{ \ln \vert \psi_N(0) \vert^2 } \sim - \lambda_1(W) N$ :
the figure shows that the slope $\lambda_1(W)$ 
 vanishes linearly $\lambda_1(W) \sim (W_c^{pool}-W)$
 (see Eq. \ref{xideloc} with $\nu_{loc}=1$).
(b)  Width $\sigma= (\overline{z^2})^{1/2}$
of the relative variable 
$z=\ln \vert \psi_N(0) \vert^2 - \overline{\ln \vert \psi_N(0) \vert^2  }$
in the limit $N \to +\infty$ as a function of the disorder strength $W$ :
it remains finite both in the localized and delocalized phase, 
but it presents a cusp
at the critical point $W_c$. }
\label{fig44deloc}
\end{figure}

In the delocalized phase, the root weight $\vert \psi_N(0) \vert^2$
presents the typical decay of Eq. \ref{Lyapunovpsi0}.
We show on Fig. \ref{fig44deloc} (a) that the corresponding
Lyapunov exponent $\lambda_1(W)$ vanishes linearly in the critical region
\begin{eqnarray}
\lambda_1(W) \opsimeq_{W \to W_c^-} (W_c-W)
\label{xideloc}
\end{eqnarray}
i.e. we find the same exponent $\nu_{loc}=1$ as
 for the divergence of the correlation length of
the localized phase (see Eq. \ref{nueq1}). 
This can be understood from the tail analysis
presented in the Appendices A and B.

We find that the width of the relative variable 
$z=\ln \vert \psi_N(0) \vert^2 - \overline{\ln \vert \psi_N(0) \vert^2  }$
remains finite in the limit $N \to +\infty$
both in the localized phase and in the delocalized phase :
its behavior as a function of the disorder strength $W$ presents
a cusp at $W_c$ as shown on Fig. \ref{fig44deloc} (b).

\begin{figure}[htbp]
 \includegraphics[height=6cm]{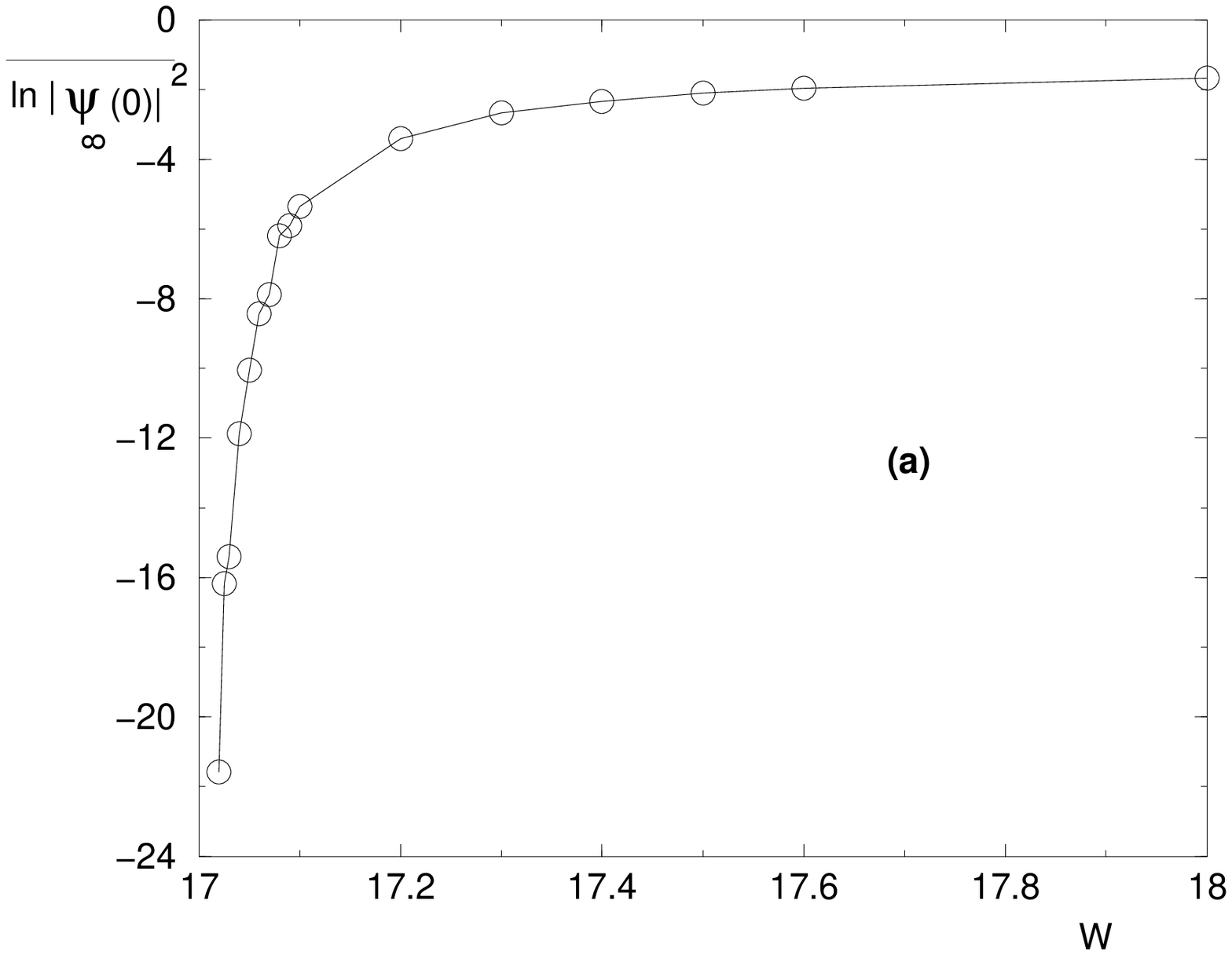}
\hspace{1cm}
\includegraphics[height=6cm]{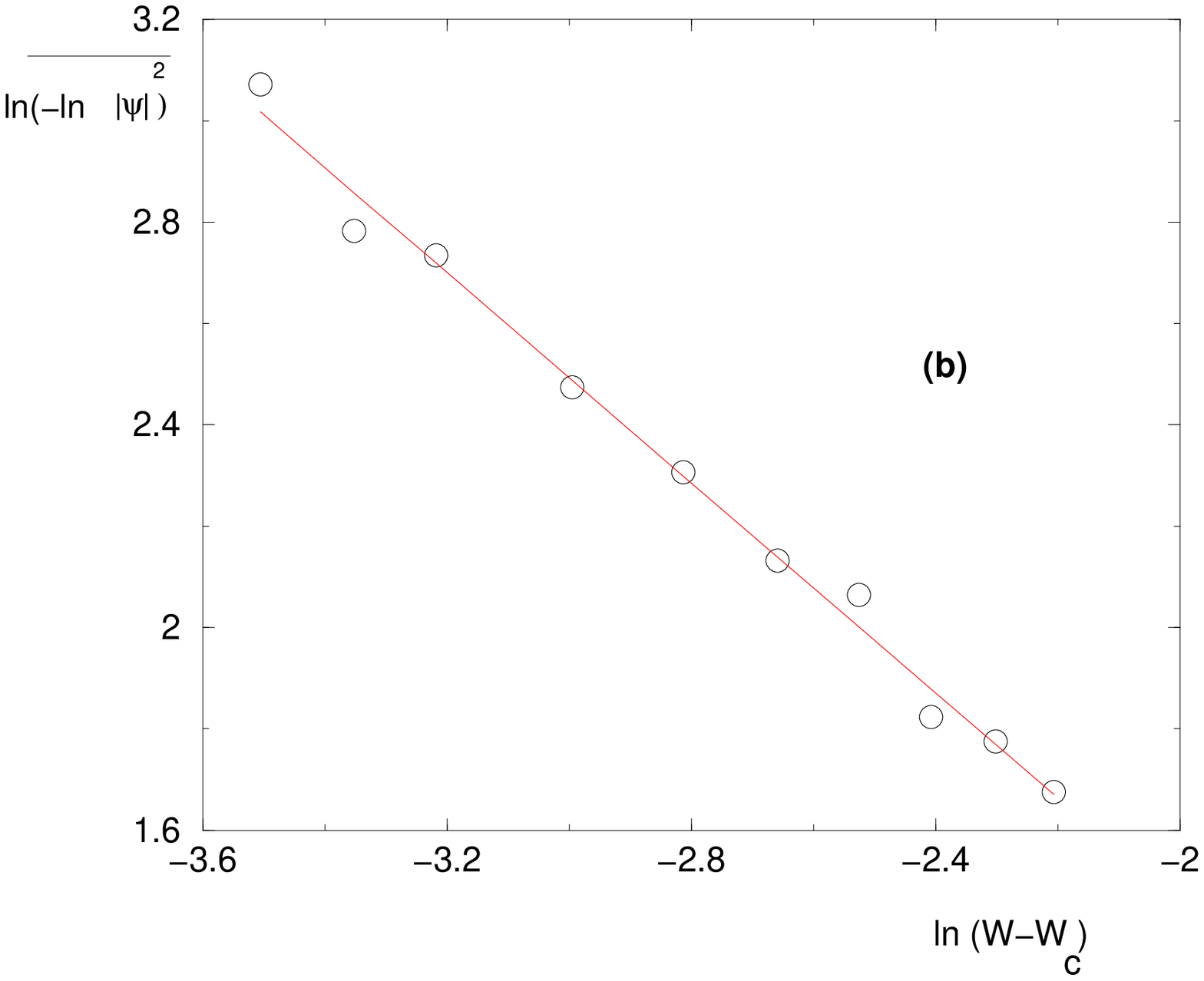}
\caption{ Critical behavior of the root weight 
$\vert \psi_{\infty}(0) \vert^2$
of an infinite tree in the localized phase :
(a)  $\overline{ \ln \vert \psi_{\infty}(0) \vert^2 }$ 
as a function of the disorder strength $W$
(b) $ \ln (- \overline{ \ln \vert \psi_{\infty}(0) \vert^2 } ) $ 
as a function of $\ln (W-W_c)$ to measure the exponent
 of the essential singularity
of Eq. \ref{essentialsingularity44loc}  :
we measure asymptotically a slope of order $\kappa_0 \sim 1 $.}
\label{fig44loc}
\end{figure}

In the localized phase, the root weight remains finite as $N \to \infty$.
As show on Fig. \ref{fig44loc}, we measure the following essential singularity
\begin{eqnarray}
 \overline{ \ln \psi_{\infty}(0) }
 \opsimeq_{W \to W_c^+} - \frac{1}{(W-W_c)^{\kappa_0}} \ \ {\rm with}
\ \ \kappa_0 \simeq 1
\label{essentialsingularity44loc}
\end{eqnarray}

As in section \ref{fsstrans}, we now discuss the
finite-size scaling in the critical region.
If there exists some finite-size scaling in the critical region for
the typical root weight, the matching of our results in the 
delocalized phase  (see Eq. \ref{Lyapunovpsi0} and \ref{xideloc}) and in the localized phase 
(Eq. \ref{essentialsingularity44loc})
requires a finite-size correlation length exponent $\nu^{FS}_{0}$ of order
\begin{eqnarray}
\nu^{FS}_{0}  =1+\kappa_{0} \simeq 2.
\label{nuFSzero}
\end{eqnarray}
We have performed the analysis described in section \ref{fsstrans}
and we measure that the relaxation length $\xi_{relax}(W)$ towards
the finite value of Eq. \ref{essentialsingularity44loc}
diverges with an exponent
\begin{eqnarray}
 \xi_{relax}(W)\oppropto \frac{1}{(W-W_c)^{\nu_{relax}}} \ \ {\rm with }
\ \ \nu_{relax} \simeq 2.
\label{taurelaxipp}
\end{eqnarray}
of the order of the exponent $\nu^{FS}_{0}$ of Eq. \ref{nuFSzero}.
We thus obtain that the critical properties are again qualitatively similar
to the critical properties described in Ref \cite{simon}
 (see the summary in Appendix C) : the traveling phase is characterized by
a velocity that vanishes linearly, but the finite size scaling is governed by
the relaxation length towards the asymptotic finite value of the non-traveling phase.
Exactly at criticality, we thus expects the following
stretched exponential decay of the typical transmission
\begin{eqnarray}
\overline{ \ln \vert \psi_{N}(0) \vert^2  } \vert_{W=W_c} \simeq -  N^{\rho_{0}}
\label{ippcriti}
\end{eqnarray}
where the exponent $\rho_{0}$ is related to the other exponents by
(see the scaling relations of Eqs \ref{kapparelation} and \ref{nutyprelation} in Appendix C)
\begin{eqnarray}
\rho_{0} = \frac{ \kappa_{0}}{\nu^{FS}_{0}} = 1- \frac{1}{\nu^{FS}_{0}}
\label{ipprho}
\end{eqnarray}
From our previous estimate of the exponent $\kappa_{0} \simeq 1.$, this corresponds to
the numerical value
\begin{eqnarray}
\rho_{0} = \frac{ \kappa_{0}}{1+\kappa_{0}} \simeq 0.5
\label{ipprhonume}
\end{eqnarray}
Again, as explained after Eq. \ref{transrhonume}, we have not been able to measure this stretched exponential behavior
 exactly at criticality from our data,
because a precise measure of the exponent $\rho_0$ would require to be exactly at the critical point.

\section{ Conclusions and perspectives }

\label{conclusion}

In summary, for the Anderson model on the Bethe lattice,
we have studied numerically the statistics of the Landauer transmission $T_N$
and the statistics of eigenstates at the center of the band $E=0$,
as a function of the disorder strength $W$ and the number $N$ of generations.
We have shown that both the localized phase and the delocalized phase
are characterized by the traveling wave propagation
 of some probability distributions.
In the text, we have presented detailed numerical results, and
in Appendices A and B, we have explained
 how the velocities of the traveling waves are determined by the tails,
via the properties of the integral kernel introduced in \cite{abou}.
The Anderson transition then corresponds to a traveling/non-traveling
critical point for these traveling waves, and the critical properties
obtained are very similar to the traveling-wave phase transition
studied in \cite{simon} :
(i) the finite value of the non-traveling phase presents an essential singularity (ii) the relaxation length towards this essential singularity determines
the finite-size scaling in the critical region
(iii) the finite-size correlation length exponent $\nu_{FS}$ is different
from the value $\nu_{loc}=1$ that govern the vanishing of the velocity
in the traveling-wave phase.
We thus hope that in the future, these properties for Anderson localization
on the Cayley tree
will be better understood by extending the methods
 that have been developed recently
for the class of the Fisher-KPP traveling waves
 (see \cite{simon,brunet} and references therein).

How results obtained on the Bethe lattice are related to 
properties of Anderson localization in finite dimension $d$
is of course a difficult question. 
During the history of localization, many values for the upper
critical dimension have been proposed such as $d_c=4,6,8,+\infty$.
Even within the supersymmetric community, 
there seems to be different interpretations.
 Mirlin and Fyodorov  \cite{dinfty} consider
 that the essential singularities
that appear on the Bethe lattice, are directly related
 to the exponential growth 
of the number of sites with the distance, 
and will become conventional power-law
behaviors as soon as $d<\infty$, so that the upper critical dimension
 is actually $d_c=+\infty$. On the contrary, Efetov \cite{efetovbook}
argues that the results obtained for the
Bethe lattice are relevant to Anderson transition in high dimension $d$
when reinterpreted within the so-called 'effective medium approximation'.
 For the Landauer transmission, the upper critical dimension $d_c$ can be
defined as the dimension where the exponent $\omega(d)$ concerning
the width of the sample-to-sample distribution in the localized phase
(see Eq.\ref{transd}) vanishes $\omega(d_c)=0$. 
This means that for $d \geq d_c$, the probability distribution
would travel as a traveling wave in the whole localized phase.
To determine whether $d_c=+\infty$ or $d_c<+\infty$,
it would be thus interesting 
to understand whether the traveling wave propagation 
 with a fixed shape is possible only on trees
or whether it can also occur in sufficiently high dimension $d$.

In low dimensions, the probability distribution 
of the logarithm of the transmission
is known to broaden with some exponent $\omega(d)>0$ 
(see Eq. \ref{transd}) in the localized phase,
whereas the probability distribution of the logarithm
 of I.P.P. is known to shrink with the system size in the delocalized phase
\cite{markosIPP} 
(see also Fig. (4a) of \cite{DP3Dmultif} where the same phenomenon occurs for 
the directed polymer in $1+3$ dimensions).
However traveling wave propagations of 
the probability distribution $P(\ln I_q)$
of Inverse Participation Ratios $I_q$ have actually been found
in finite dimensions but only {\it exactly at criticality}.
This phenomenon has been first obtained 
for the power-law random band matrix model \cite{Mirlin_Evers}
and has been then observed for the Anderson model
 in dimensions $d=3,4$ \cite{Mirlin2002} :
the motions of the typical values
 $\overline{ \ln I_q}$ with $(\ln L)$ determine
the multifractal spectrum, whereas the reduced variables
 $y=I_q/I_q^{typ}$ keep fixed distributions presenting 
 power-law tails with $q$-dependent exponents.
We have actually observed the same behavior 
for the directed polymer in $1+3$ dimensions
exactly at the localization/delocalization transition
(see Fig. (3a) and (3b) of Ref. \cite{DP3Dmultif}).
In all these cases, it would be thus very nice to better understand 
the relations between the tails of the distributions
 and the motions of the typical values, and to identify the wave equation
that underlies these traveling waves.

Finally, our work raises once again the question whether
Anderson transitions are characterized by 
several length scales that diverge with various $\nu$ exponents.
Here for the Cayley tree, besides the exponent $\nu_{loc}=1$
that has been predicted for a long time \cite{Kun_Sou}, we have found 
other diverging lengths with the following physical meanings.
(i) in the localized phase, besides the exponent $\nu_{loc}=1$
associated to the divergence of localization length
$\xi_{loc} \sim 1/(W -W_c)^{\nu_{loc}=1}$ that describes the far exponential
decay of the transmission $T$ (Eq. \ref{transloc}),
we have found that the critical behavior of the entropy
is governed by another diverging length scale with $\nu_s \sim 1.5$
(Eq. \ref{xiSloc}) that characterizes 
 the size where the weight $\vert \psi(x) \vert^2$ is concentrated.
(ii) in the delocalized phase, we have found that the Landauer transmission
reaches its asymptotic value after a length diverging as $\nu_T^{FS} \sim
1.25$ (Eq. \ref{nuFStrans})
that governs the finite-size scaling of the transmission in the
critical region. Even if the precision of numerically measured 
critical exponents can be always discussed, 
we feel nevertheless that our numerical results
are not compatible with the existence of the single exponent $\nu_{loc}=1$.
Note that for the directed polymer on the Cayley tree, we have also found
previously that the critical properties
 involve two exponents $\nu=2$ and $\nu'=1$ \cite{DPcayleycriti}.
Whether there exists various exponents $\nu$ 
for Anderson model in finite dimension $d$
has been debated for a long time. The majority of papers on Anderson
localization seem in favor of a single $\nu$ (see the reviews 
\cite{thouless,souillard,bookpastur,Kramer,markos,mirlinrevue}),
but it seems to us that numerical papers actually study always the same type
of observables. In particular, we have not been able to find
numerical studies concerning the entropy of eigenstates. 
On the theoretical side, we are not aware of many theories
in favor of various $\nu$ exponents, except
(i)  the supersymmetric studies of Refs \cite{efetov2lengths,efetovbook}
that involve two different diverging length scales, 
called respectively the localization length and the phase coherence length.
(ii) the pseudo-delocalization transition of the random hopping model
in dimension $d=1$, that is characterized by 
the two exactly known correlation lengths exponents
$\nu=1$ and $\nu=2$ (see \cite{loc1d} and references therein), 
as in the other models
described by the same strong disorder fixed point 
(see the review \cite{SDreview}).
We feel that if localization models exhibit several $\nu$ exponents
both on the Cayley tree which represents some mean-field $d=\infty$ limit
and in some models in dimension $d=1$, the possibility of various $\nu$
for Anderson transition in dimension $d=3$ should be reconsidered
by studying in detail the statistics of various observables.

\appendix

\section{ Tail analysis for the traveling wave
of the Landauer transmission in the localized phase }

In this Appendix, we translate the analysis of Ref. \cite{abou}
concerning the distribution of the self-energy in the localized phase
for the traveling-wave propagation of the Landauer transmission
discussed in section \ref{trans}.

The recursion of Eq. \ref{riccatiinside} for the complex Riccati variable $R_n$
reads more explicitly in terms of its real and imaginary parts $R_n=X_n-iY_n$ with $X_n \in ]-\infty,+\infty[$ and
$Y_n \in [0,+\infty[$.

\begin{eqnarray}
X_n && = - \epsilon_n - \sum_{m=1}^K \frac{X_{n-1}(m)}{ X_{n-1}^2(m)+Y_{n-1}^2(m) } \\
Y_n && = \sum_{m=1}^K \frac{Y_{n-1}(m)}{ X_{n-1}^2(m)+Y_{n-1}^2(m) }
\label{ricattiinsideXY}
\end{eqnarray}
where the random on-site energies $\epsilon_n$ are drawn from the flat distribution 
of Eq. \ref{flat}. The Landauer transmission of Eq. \ref{transresume} then reads
\begin{eqnarray}
T_n= 1 - \left \vert \frac{X_n +i (1-Y_n)}{-X_n+i(1+Y_n)} \right \vert^2 = 
\frac{ 4 Y_n }{ X_n^2 + (Y_n+1)^2}
\label{transapp}
\end{eqnarray}

In the delocalized phase, this recursion of Eq. \ref{ricattiinsideXY}
 is difficult to analyze because
one has to find the joint distribution of the two finite variables $(X,Y)$
that remain stable upon iteration. In the localized phase however, 
the problem is simpler  \cite{abou} as we now recall.

\subsection{  Linearized recursions in the localized phase }

In the localized phase, if the imaginary part 
$Y_n$ converge towards zero exponentially in $n$ as 
\begin{eqnarray}
Y_n  =  e^{- v n} y_n
\label{ynscaling}
\end{eqnarray}
where $v>0$
and where $y_n$ remains a finite random variable upon iteration, 
 one has then to study the simpler recurrence for large $n$  \cite{abou,MD}
\begin{eqnarray}
X_n && = - \epsilon_n - \sum_{m=1}^K \frac{1}{ X_{n-1}(m) } \nonumber  \\
e^{-v} y_n && = \sum_{m=1}^K \frac{y_{n-1}(m)}{ X_{n-1}^2(m)}
\label{ricattilinearized}
\end{eqnarray}
Note that the form of Eq. \ref{ynscaling} corresponds to a traveling wave
of velocity $v$ for the variable $(\ln Y_n)$
\begin{eqnarray}
\ln Y_n  =  - v n  + \ln y_n
\label{lnynscaling}
\end{eqnarray}

The real part $X_n$ satisfies now a closed recurrence independent of the $Y_n$,
and its stable distribution $P^*(X)$ satisfies the closed equation
\begin{eqnarray}
P^*(X)= \int d\epsilon p(\epsilon) \int dX_1 P^*(X_1) ...\int dX_m P^*(X_m) 
\delta \left[ X +\epsilon +  \sum_{m=1}^K \frac{1}{ X_m } \right]
\label{petoileX}
\end{eqnarray}
An important property of this distribution
 is that it presents the power-law tail
\begin{eqnarray}
P^*(X) \opsimeq_{ X \to \pm \infty}  \frac{ K P^*(0) }{ X^2 }
\label{tailX}
\end{eqnarray}
because whenever one of the $K$ variables $X_i$ on the right handside
of Eq. \ref{petoileX} is close to $0$, the variable $X$ of the left-handside
is large with $X \sim - 1/X_i$.

The stable joint distribution $P^*(X,y)$ satisfies
\begin{eqnarray}
P^*(X,y)= \int d\epsilon p(\epsilon) \int dX_1 dy_1 P^*(X_1,y_1)
 ...\int dX_m dy_m P^*(X_m,y_m) 
\delta \left[ X +\epsilon +  \sum_{m=1}^K \frac{1}{ X_m } \right]
\delta \left[ y - \sum_{m=1}^K \frac{ y_m} { e^{-v} X_m^2 } \right]
\label{ricattistab}
\end{eqnarray}

\subsection{  Tail analysis  }

The idea of \cite{abou} is to look for the power-law tail in $y$ that is compatible 
with the recursion equation,
i.e. one assumes
\begin{eqnarray}
P^*(X,y) \opsimeq_{y \to \infty} \frac{ A(X)}{y^{1+\beta} }
\label{tail}
\end{eqnarray}
with some exponent $0<\beta<1$.
Note that, in the traveling wave language of Eq. 
\ref{lnynscaling}, this is equivalent to look for
solutions with an exponential tail $e^{-\beta u}$
for the variable $u=\ln Y_n +  n v=\ln y_n$.
So this corresponds to the usual exponential tail analysis of fronts 
\cite{revues_traveling,brunetreview}, except for the following difference :
in usual studies of propagation into unstable phases, it is the 'forward tail'
that has an exponential decay and that determines the velocity,
whereas in our present case, it is the 'backward tail' that determines the
propagation.

In Laplace transform with respect to $y$, 
the power-law decay of Eq. \ref{tail}
 corresponds to 
the following singular expansion near the origin
\begin{eqnarray}
{\hat P}^*(X;s) \equiv \int_0^{+\infty} dy e^{-s y}
P^*(X,y) && = P^*(X) - \int_0^{+\infty} dy (1-e^{-s y}) P^*(X,y)
= P^*(X) - \int_0^{+\infty} \frac{dv}{s} (1-e^{- v}) P^*(X,\frac{v}{s}) 
\nonumber \\
&& \opsimeq_{s \to 0}   P^*(X) - s^{\beta} A(X) \int_0^{+\infty}
 dv \frac{(1-e^{- v})}{v^{1+\beta}}
\label{devlaplace}
\end{eqnarray}
Equation \ref{ricattistab}
 becomes for the Laplace transform ${\hat P}^*(X;s)$
\begin{eqnarray}
{\hat P}^* (X;s) =  \int d\epsilon p(\epsilon) \int dX_1 {\hat P}^*(X_1,
\frac{s}{\lambda X_1^2}) 
...\int dX_m  {\hat P}^*(X_m,\frac{s}{\lambda X_m^2}) 
\delta \left[ X +\epsilon +  \sum_{m=1}^K \frac{1}{ X_m } \right]
\label{ricattistablaplace}
\end{eqnarray}
Using the expansion of Eq. \ref{devlaplace}, one obtains
 that the function $A(X)$
has to satisfy the eigenvalue equation 
\begin{eqnarray}
e^{- v \beta} A(X) = K \int 
\frac{dX_1}{(X_1^2)^{\beta}}   Q(X+\frac{1}{X_1}) A(X_1)
\label{eqpropreAX}
\end{eqnarray}
where the function $Q(u)$ represents the stationary distribution
of the variable  $u=-\epsilon -  \sum_{m=2}^K \frac{1}{ X_m }$
\begin{eqnarray}
Q(u) = 
   \int d\epsilon p(\epsilon) 
 \int dX_2 P^*(X_2)
...\int dX_m   P^*(X_m) 
\delta \left[ u +\epsilon +  \sum_{m=2}^K \frac{1}{ X_m } \right]
\label{fonctionQ}
\end{eqnarray}
(see Eqs 6.5 and 6.6 in \cite{abou}).

\subsection{  Eigenvalue problem for an integral kernel }

The tail analysis thus leads to the  eigenvalue problem the integral kernel appearing
in Eq. \ref{eqpropreAX}
\begin{eqnarray}
\Lambda A(X) = K \int 
\frac{dX_1}{(X_1^2)^{\beta}}   Q(X+\frac{1}{X_1}) A(X_1)
\label{kernel}
\end{eqnarray}
For each $\beta$ and $W$, the integral kernel is positive,
and thus one expects some continuous analog of the Perron-Froebenius theorem :
the iteration of the integral kernel will
 converge towards a positive eigenvector $A_0(X)$
that is associated to the maximal eigenvalue $\Lambda_0(\beta,W)$ of the kernel.
For instance, for the special case $\beta=0$, the solution is simply $A(X)=P^*(X)$
(see Eqs \ref{petoileX} and \ref{fonctionQ}) and 
\begin{eqnarray}
\Lambda_0(\beta=0,W)=K
\label{lambda0beta0}
\end{eqnarray}
Except for this case $\beta=0$, we are not aware of any explicit solution
for the eigenvalue  $\Lambda_0(\beta,W)$ 
(even for the simpler case where the distribution
of the on-site energies $p(\epsilon)$ is a Cauchy
 law and where the distributions $P^(X)$
and $Q(u)$ are also Cauchy laws \cite{abou,MD}).
With Eqs \ref{eqpropreAX} and \ref{kernel},
one concludes that at fixed $W$,
each mode $\beta$ is associated to the
 velocity $v(\beta,W)$ ( Eq. \ref{ynscaling}) 
\begin{eqnarray}
v (\beta,W) = - \frac{1}{\beta} \ln  \Lambda_0(\beta,W) 
\label{vbeta}
\end{eqnarray}

\subsection{  Selection of the tail exponent
 and of the velocity of the traveling wave }

The selection of the tail exponent $\beta$ of Eq. \ref{tail} 
and of the corresponding velocity $v(\beta)$
of Eq \ref{ynscaling} usually depend on the form of the initial condition
\cite{Der_Spohn,revues_traveling,brunetreview}. 
In our present case, the initial condition
  is completely localized (see Eq. \ref{riccatifirst})
\begin{eqnarray}
P_{ini}(Y) = \delta(Y-1)
\label{initialtraveling}
\end{eqnarray}
In this case, one expects that the solution that will
be dynamically selected \cite{Der_Spohn,revues_traveling,brunetreview}
corresponds to the tail exponent $\beta_{selec}(W)$
and to the velocity $v_{selec}(W)=v(\beta_{selec}(W),W)$
determined by the following extremization
\begin{eqnarray}
0= \left[ \partial_{\beta} v(\beta,W) \right]_{\beta=\beta_{selec}(W)}
= \left[ \frac{1}{\beta^2} \ln  \Lambda_0(\beta,W) 
- \frac{1}{\beta} \frac{ \partial_{\beta}\Lambda_0(\beta,W) }
{\Lambda_0(\beta,W)}
\right]_{\beta=\beta_{selec}(w)}
\label{selection}
\end{eqnarray}
The critical point is then determined by the two conditions
\begin{eqnarray}
v (\beta_c,W_c) && =0 \\
\partial_{\beta_c} v (\beta_c,W_c) && =0 
\label{criticonditions}
\end{eqnarray}

\subsection{ Example in the 'strong disorder approximation' }

Since the general discussion is rather obscured by the absence
 of an explicit expression
for the eigenvalue $\Lambda_0(\beta,W)$ of the kernel of Eq. \ref{kernel},
it is useful to consider the following strong disorder approximation
 (called 'upper limit' condition in \cite{abou}),
where the recursion for the $X_n$ in Eq. \ref{ricattilinearized}
is simply replaced by \cite{abou}
\begin{eqnarray}
X_n \simeq \epsilon_n
\label{strongapprox}
\end{eqnarray}
The argument is that in the limit of very large $W$,
the distribution of $X$ has also a width of order $W$, 
so that the neglected terms in
Eq. \ref{ricattilinearized} are of order $K/W$. 
Of course the approximation of Eq. \ref{strongapprox} 
is not very well controlled, because
it concerns random variables, and it suppresses
 important correlations between the variables
$(X,Y)$ of Eq. \ref{ricattilinearized}. Nevertheless, it is useful to consider it
before returning to the true recursions of Eq. \ref{ricattilinearized}, because
with Eq. \ref{strongapprox}, the eigenvalue $\Lambda_0(\beta,W)$
is replaced by simple expression \cite{abou}
(see also section 7.2 of \cite{MD} where the approximation of independence
between $(X,Y)$ is considered for the Cauchy case)
\begin{eqnarray}
\Lambda_0^{SD} (\beta,W) =
 K \int d\epsilon p(\epsilon) \vert \epsilon \vert^{-2 \beta} 
\label{eigenapprox}
\end{eqnarray}
The velocity then reads
\begin{eqnarray}
v^{SD} (\beta,W) = - \frac{1}{\beta}
 \ln \left( K \int d\epsilon p(\epsilon) \vert \epsilon \vert^{-2 \beta} \right)
\label{vapprox}
\end{eqnarray}
For the flat distribution
 $p(\epsilon)$ of random site energies (Eq. \ref{flat}),
one thus obtains
\begin{eqnarray}
v^{SD} (\beta,W) = - \frac{1}{\beta}
 \ln \left( \frac{K}{1-2 \beta} \left( \frac{2}{W} \right)^{2 \beta} \right)
= 2 \ln \frac{W}{2} - \frac{1}{\beta}
 \ln \left( \frac{K}{1-2 \beta} \right)
\label{vapproxflat}
\end{eqnarray}
The function $v^{SD} (\beta)$ is defined on the interval $0<\beta<1/2$ :
 it flows towards $(-\infty)$ in the limits 
$\beta \to 0$ and $\beta \to 1/2$
with the following behaviors
\begin{eqnarray}
v^{SD} (\beta,W) \opsimeq_{\beta \to 0} - \frac{ \ln K} {\beta}
\label{vapproxflatzero}
\end{eqnarray}
and
\begin{eqnarray}
v^{SD} (\beta,W) \opsimeq_{\beta \to 1/2}
 - \frac{1} {\beta} \ln \frac{1}{1-2 \beta}
\label{vapproxflatdemi}
\end{eqnarray}
It has a single maximum at $\beta_{selec}^{SD}$ 
given by the extremum condition of Eq. \ref{selection},
so that $\beta_{selec}^{SD}$ is actually independent of $W$.
For $K=2$, the selected exponent is of order $\beta_{selec}^{SD}
 \simeq 0.3133 $ for any $W$.
We note that is value is close to the value $\beta(W=100) \simeq 0.33$
that we measure for the large disorder $W=100$ (see Figure \ref{figtranswidth} a).
In conclusion, this 'strong disorder approximation' allows to see
explicitly how things work on a simple example, 
and seems to give a reasonable value of
$\beta_{selec}$ for very large $W$.

\subsection{  Argument in favor of $0<\beta_{selec}(W) < 1/2$ }

\label{beta0demi}

We now return to the analysis of the full problem of Eq. \ref{ricattilinearized}.
Here we should say that we do not agree with the discussion of \cite{abou}
concerning the selection of the tail exponent $\beta$ :

(i) in \cite{abou}, the authors
 conclude that within the localized phase, $\beta_{selec}(W)$ 
decreases from unity and reaches $\beta_c=1/2$ exactly at criticality
(see the text between Eqs. 6.8 and 6.9 in \cite{abou}).

(ii)  in our numerical results of section \ref{trans} 
 we have found instead that the selected exponent 
$\beta_{selec}(W)$ is always smaller than $1/2$ :
it slightly grows as the disorder strength $W$ decreases
and has a value close to $0.5$ near criticality 
(see Figure \ref{figtranswidth} a).

\begin{figure}[htbp]
 \includegraphics[height=6cm]{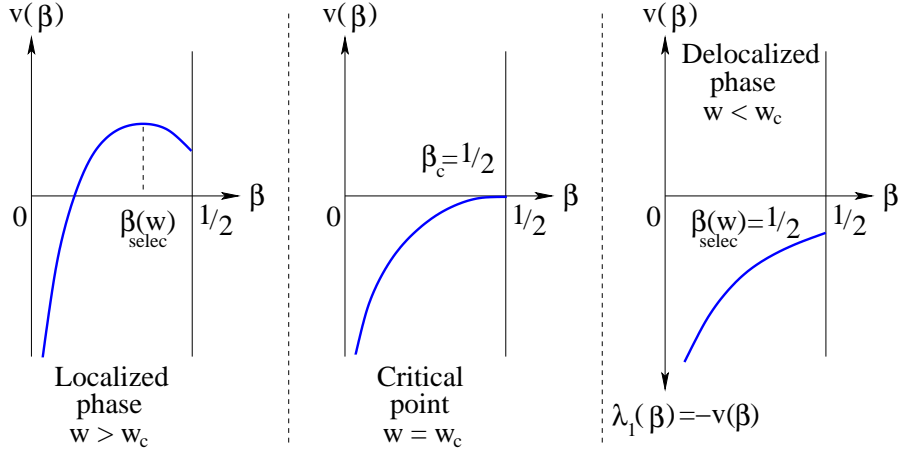}
\caption{ Shape of the velocity $v(\beta,W)$ as a function of the tail
exponent $\beta$ for various disorder strength $W$ :
(i) in the localized phase $W>W_c$, 
the velocity is extremum at a $W$-dependent value $\beta_{selec}(W)<1/2$
and it is positive $v_{selec}>0$ 
(ii) at the critical point $W=W_c$, the velocity is extremum at
$\beta_c=1/2$ where it vanishes $v_c=0$
(iii) in the delocalized phase, there exists another traveling-wave 
moving in the other direction of velocity $\lambda_1(\beta,W)=-v(\beta,W)>0$ :
this velocity is extremum at $\beta=1/2$ for any $W<W_c$
(see Appendix B for more details) }
\label{figvelocity}
\end{figure}

We propose the following argument to justify our
 finding $0<\beta_{selec}(W) < 1/2$.
We think that the tail analysis is actually well defined
only on the interval $0<\beta<1/2$ for the following reasons.
 The eigenvalue Eq. \ref{kernel}
relates the behavior of $A(X)$ at $\vert X \vert \to \infty$
to the behavior of $A(X)$ near the origin $X \to 0$.
For $X \to \infty$, the function $Q(X+\frac{1}{X_1})$ will be finite
 in the region where $( X+\frac{1}{X_1})$ is finite,
i.e. the integration is dominated by the region $X_1 \sim -1/X$
and we obtain the power-law decay
\begin{eqnarray}
 A(X) \opsimeq_{ \vert X \vert \to \infty }
 \frac{K A(0)}{ \Lambda_0(\beta) \vert X \vert^{2-2 \beta} } 
 \label{AXinfinity}
\end{eqnarray}
However the function $A(X)$ has to be integrable at $ \vert X \vert \to \infty$
to obtain from Eq. \ref{tail} the probability of the variable $y$
alone at large $y$
\begin{eqnarray}
P^*(y) = \int_{-\infty}^{+\infty} dX P^*(X,y) 
\opsimeq_{y \to \infty} \frac{ \int_{-\infty}^{+\infty} dX A(X)}{y^{1+\beta} }
\label{tailinteg}
\end{eqnarray}
According to Eq. \ref{AXinfinity}, 
the function $A(X)$ is integrable only for
\begin{eqnarray}
0<\beta< \frac{1}{2}
\label{betacondition}
\end{eqnarray}
i.e. we obtain that the tail analysis has a meaning only for 
$0 \leq \beta<1/2$. 
In the limit $\beta \to 0$, 
we expect from Eq. \ref{lambda0beta0} the same behavior as 
in Eq. \ref{vapproxflatzero}
\begin{eqnarray}
v (\beta,W) \opsimeq_{\beta \to 0} - \frac{ \ln K} {\beta}
\label{vbetazero}
\end{eqnarray}
Exactly at $\beta=1/2$, the authors of \cite{abou} have argued that 
\begin{eqnarray}
\left[ \partial_{\beta} \Lambda_0 (\beta,W) \right]_{\beta=1/2} =0 
\label{partialdemi}
\end{eqnarray}
for any $W$, as a consequence of the symmetry 
$\Lambda_0 (\beta,W)=\Lambda_0 (1-\beta,W)$ coming from the consideration
of the adjoint kernel
 (see more details around Eqs. 6.7 and 6.8 in \cite{abou}).
Here in contrast to \cite{abou}, we think that the region $\beta>1/2$
is not physical because of Eq. \ref{AXinfinity}, but the condition of Eq. 
\ref{partialdemi} is useful to understand
 why the critical point
determined by Eqs \ref{criticonditions} corresponds to the tail exponent
\cite{abou}
\begin{eqnarray}
\beta(W ) \opsimeq_{W \to W_c^+}  \beta_c=\frac{1}{2} 
\label{betacdemi}
\end{eqnarray}
To summarize the selection mechanism, 
the shape of the velocity $v(\beta,W)$ as a function of the tail
exponent $\beta$ in shown on Fig. \ref{figvelocity}
for various disorder strength $W$ :

(i) in the localized phase $W>W_c$, the velocity derivative 
is negative at $\beta=1/2$ : 
\begin{eqnarray}
\left[ \partial_{\beta} v(\beta,W>W_c) \right]_{\beta=1/2}
= \left[ \frac{1}{\beta^2} \ln  \Lambda_0(\beta,W>W_c)
 \right]_{\beta=1/2}
= - \left[ \frac{1}{\beta} v(\beta,W>W_c) \right]_{\beta=1/2} <0
\label{derineg}
\end{eqnarray}
The velocity is extremum at a $W$-dependent value $\beta_{selec}(W)<1/2$
and the corresponding selected velocity is positive $v_{selec}>0$.

(ii) at the critical point $W=W_c$, the velocity is extremum at
$\beta_c=1/2$ where it vanishes $v_c=0$.

(iii) in the delocalized phase, the velocity derivative 
is positive at $\beta=1/2$ : 
\begin{eqnarray}
\left[ \partial_{\beta} v(\beta,W<W_c) \right]_{\beta=1/2}
= \left[ \frac{1}{\beta^2} \ln  \Lambda_0(\beta,W<W_c)
 \right]_{\beta=1/2}
= - \left[ \frac{1}{\beta} v(\beta,W<W_c) \right]_{\beta=1/2} >0
\label{deripos}
\end{eqnarray}
We explain in Appendix B that in this delocalized phase,
there exists another traveling-wave 
moving in the other direction with the velocity 
 $\lambda_1(\beta,W)=-v(\beta,W)>0$ :
this velocity $\lambda_1(\beta,W)$ is then 
extremum at $\beta=1/2$ for any $W<W_c$
(see Appendix B for more details).

As a final remark, we believe that the solution exactly at 
$W_c$ is not valid anymore because of Eq. \ref{AXinfinity},
and that the appropriate treatment exactly at criticality
should replace the finite velocity motion in $v n$ assumed in Eq.
\ref{lnynscaling} by the form
\begin{eqnarray}
\ln Y_n  =  - (cst) n^{\rho}  + \ln y_n
\label{lnynscalingcriti}
\end{eqnarray}
where the anomalous exponent $0<\rho<1$ has been discussed in the text
(see Eqs \ref{transfss} and \ref{transrhonume}).

\subsection{ Conclusion }

In conclusion, the analysis of \cite{abou}
 is thus very close to the traveling wave analysis
of the directed polymer model \cite{Der_Spohn}: the difference is that
in the directed polymer, the variables $X_n$ are random variables independent 
of the $Y_n$  and the selected $\beta_{selec}$ can be simply obtained from
the non-integer moments of $X$ (see the section 7.2 in \cite{MD}), 
whereas in the present localization model,
the variables $X_n$ and $Y_n$ are correlated and one has thus to solve 
the eigenvalue problem for the integral kernel of Eq. \ref{kernel}.
In the following Appendix, 
we explain how the same ideas can be used in the delocalized phase.

\section{ Tail analysis for the traveling waves
in the delocalized phase }

As explained in section \ref{eigen}, in the delocalized phase, we
have considered the recurrence for the real Riccati variable $X_n$
\begin{eqnarray}
X_n = - \epsilon_n - \sum_{m=1}^K \frac{1}{ X_{n-1}(m) } 
\label{ricattiX}
\end{eqnarray}
together with the recurrences of
 Eqs \ref{reccq} for the auxiliary variables $C^{(q)}$
introduced in Eq. \ref{defcq}.

\subsection{ Tail analysis leading to the same integral kernel as in Appendix A }

To make more visible the similarities with the previous Appendix,
it is convenient to perform the change of variables
\begin{eqnarray}
D^{(q)}_n  \equiv (X_n^2)^q C^{(q)}_n
\label{CD}
\end{eqnarray}
to obtain the following recursions for these variables
\begin{eqnarray}
D^{(q)}_n  =  1+ 
 \sum_{m=1}^{K} \frac{D^{(q)}_{n-1}(m)}{(X_{n-1}^2(m))^q}  
\label{recdq}
\end{eqnarray}

The recursion of Eq. \ref{ricattiX} for 
$X_n$ alone will as before converge to some
distribution $P^*(X)$ satisfying Eq. \ref{petoileX}.
in the delocalized phase, one expects that the variables $C^{(q)}$ will grow exponentially
(see Eqs \ref{LyapunovCq}). We thus set
\begin{eqnarray}
D^{(q)}_n  =  e^{n \lambda_q}  d^{(q)}
\label{scalingDn}
\end{eqnarray}
where $\lambda_q>0$ governs the exponential growth and where $d^{(q)}$
remains a finite random variable upon the iteration
\begin{eqnarray}
e^{ \lambda_q} d^{(q)}  =
 \sum_{m=1}^{K} \frac{d^{(q)}(m)}{(X^2(m))^q}  
\label{eigendq}
\end{eqnarray}
The similarity with the discussions of Appendix A is now obvious 
(see Eqs \ref{ynscaling} and \ref{ricattilinearized}).
Assuming the power-law decay
\begin{eqnarray}
P^*(X,d) \opsimeq_{d \to \infty} \frac{ \Phi(X)}{d^{1+\mu} }
\label{Dtail}
\end{eqnarray}
with $0<\mu<1$,
one finds that the function $\Phi(X)$ has to satisfy the eigenvalue problem
\begin{eqnarray}
e^{ \mu \lambda_q}   \Phi(X) = K
 \int  \frac{dX_1}{(X_1^2)^{q \mu}} Q(X+\frac{1}{X_1}) \Phi(X_1)
\label{eqproprephiX}
\end{eqnarray}
i.e. it is the same integral kernel as in Eq. \ref{eqpropreAX}
with the correspondences $\beta \to q \mu$ and 
$e^{-v \beta} \to e^{ \mu \lambda_q}$,
but now we are interested in the phase $\lambda_q>0$.

\subsection{ Selection of tail exponents }

We have argued above (see section \ref{beta0demi}) that
 the problem for the function $\Phi(X)$ is well defined only for $0< \beta=q \mu <1/2$,
i.e. the exponent $\mu_q$ that governs the power-law of Eq. \ref{Dtail}
is restricted to the interval
\begin{eqnarray}
0< \mu_q < \frac{1}{2 q}
\label{intervalmuq}
\end{eqnarray}

\begin{figure}[htbp]
 \includegraphics[height=6cm]{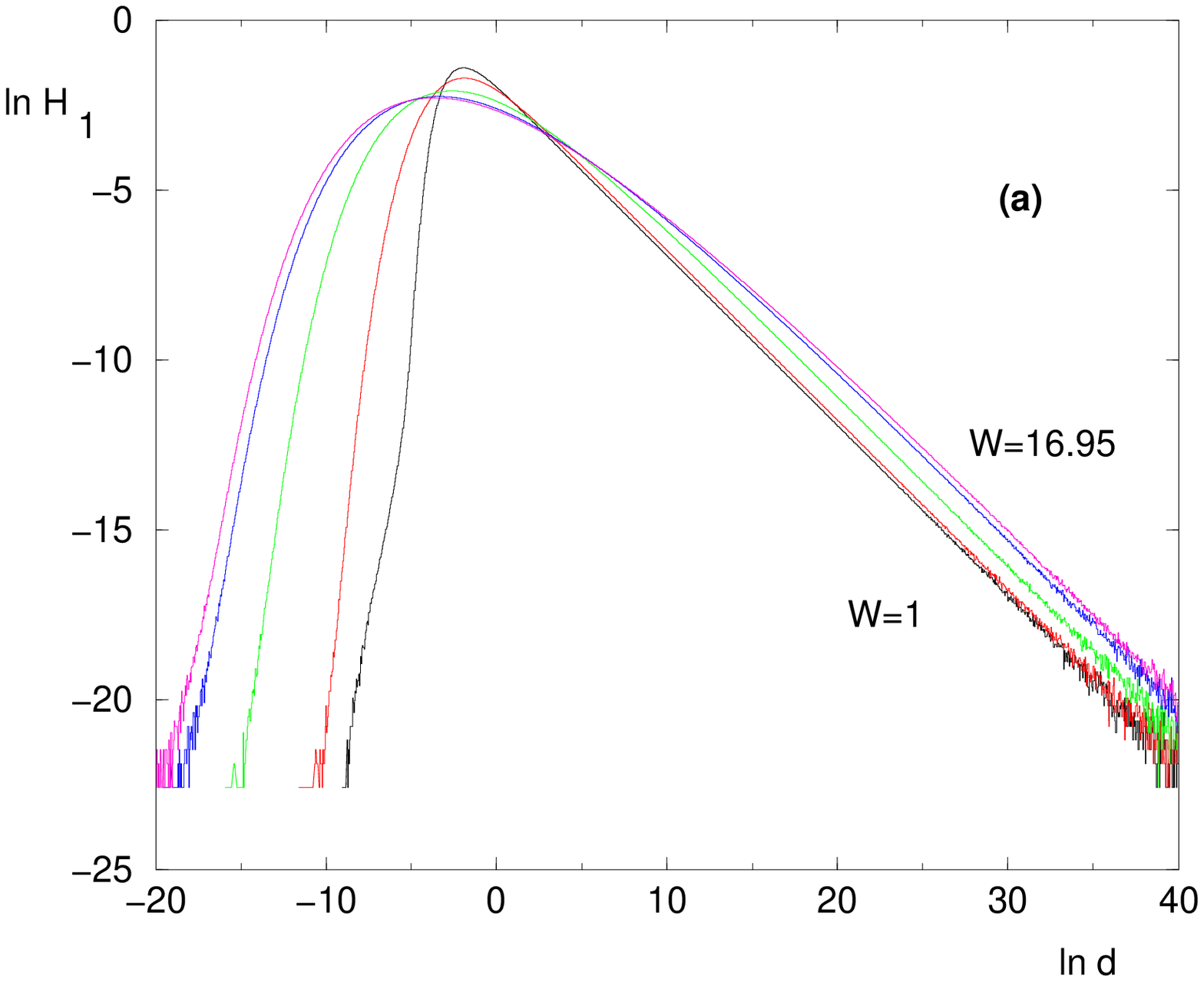}
\hspace{2cm}
\includegraphics[height=6cm]{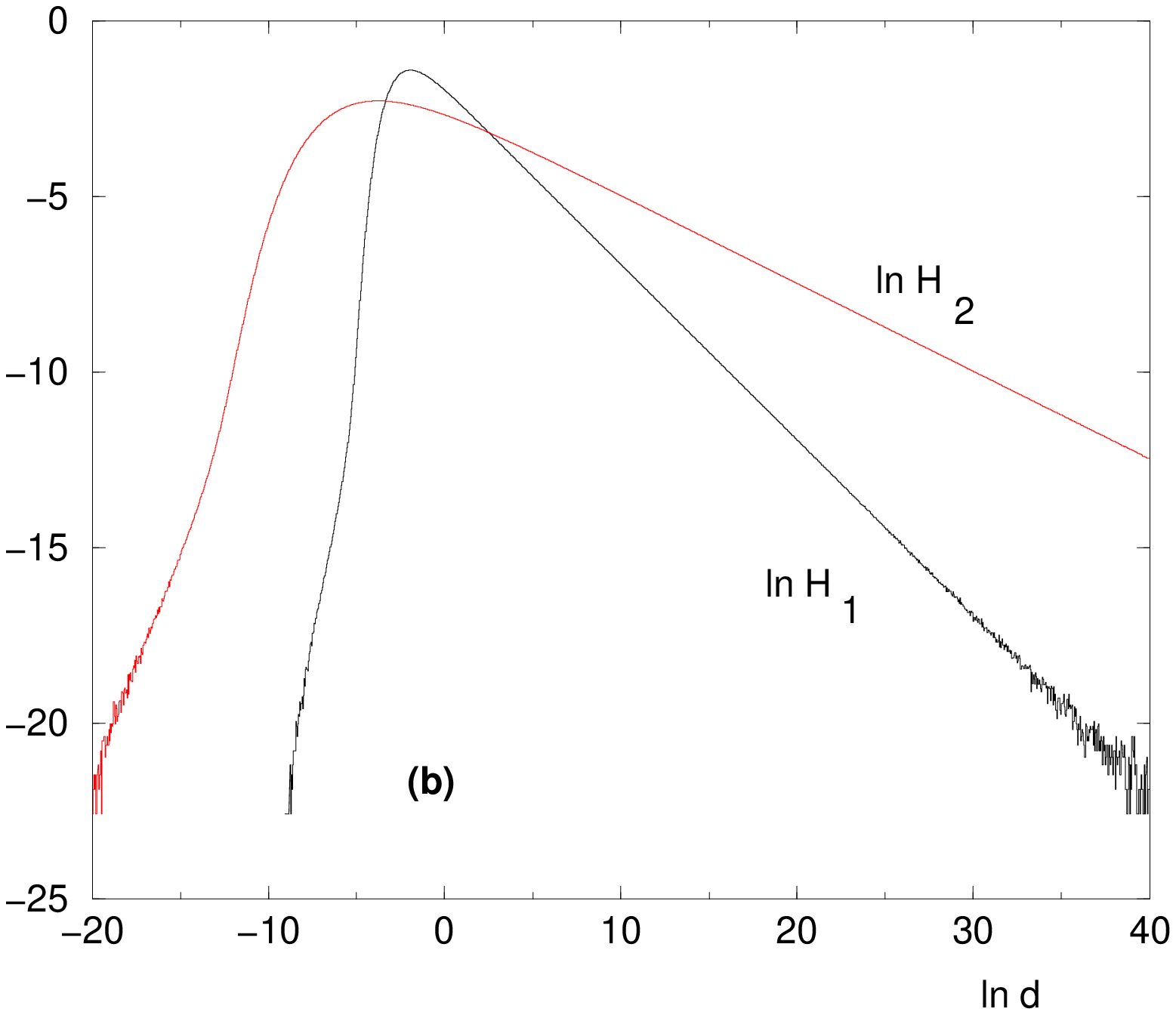}
\caption{ (a) Probability distribution $H_1$ of the variable $\ln d^{(q=1)}$ of Eq. \ref{scalingDn}
for various disorder strength $W=1,5,10,15,16.95$ :
the exponent $ \mu_1^{selec}$ of Eq. \ref{Dtail} is of order $\mu_1^{selec} \sim 0.5$
in the whole delocalized phase.
(b) For $W=1$, comparison of the probability distributions $H_1$ and $H_2$ 
of the variable $\ln d^{(q)}$ for $q=1$ and $q=2$ :
the exponents of Eq. \ref{Dtail} are respectively   $\mu_1^{selec} \sim 0.5$
and $\mu_2^{selec} \sim 0.25$ in agreement with Eq. \ref{muselec}.}
\label{fighistoc1c2}
\end{figure}

Denoting as before $\Lambda_0(\beta,W)$ the maximal eigenvalue
 of the kernel of Eq. \ref{kernel}, 
each mode $\mu$ is associated to the Lyapunov exponent (see Eq. \ref{ynscaling})
\begin{eqnarray}
\lambda_q(\mu)  = \frac{1}{\mu} \ln  \Lambda_0( q \mu,W ) 
\label{lambdaqmu}
\end{eqnarray}
From Eq. \ref{lambda0beta0}, we have the following behavior for $\mu \to 0$
\begin{eqnarray}
\lambda_q(\mu) \opsimeq_{\mu \to 0} \frac{1}{\mu} \ln K
\label{lambdaqmubeta0}
\end{eqnarray}
Near the other boundary $\mu \to 1/(2q)$, 
the property of Eq. \ref{partialdemi},
yields that the partial derivative with respect to $\mu$
is still negative at $\mu \to 1/(2q)$
\begin{eqnarray}
\left[ \partial_{\mu} \lambda_q(\mu) \right]_{\mu \to 1/(2q)} =
\left[ -\frac{1}{\mu^2} \ln  \Lambda_0( 1/2,W )  \right]_{\mu \to 1/(2q)} <0
\label{derimu}
\end{eqnarray}
For the special case $q=1$, the curve $\lambda_1(\beta)$ 
is directly related via  $\lambda_1(\beta)=-v(\beta)$
 to the curve $v(\beta)$ discussed in Appendix A and shown
on Fig. \ref{figvelocity} (see the delocalized case $W<W_c$).
Since the selected tail exponent $\mu_{selec}$ has to extremize 
$\lambda_q(\mu)$, we conclude that the selected tail exponents
are given by the simple values
\begin{eqnarray}
\mu_q^{selec} = \left( \frac{1}{2q} \right)^-
\label{muselec}
\end{eqnarray}
in the whole delocalized phase. 
This is in agreement with our numerical results.
We show on Fig. \ref{fighistoc1c2} (a) the probability distribution of
$H_{q=1}(\ln d)$ of the variable $(\ln d)$ of Eq. \ref{scalingDn}
for various disorder strength $W=1,5,10,15,16.95$ :
the tail exponent $ \mu_1^{selec}$ of Eq. \ref{Dtail} remains
of order $\mu_1^{selec} \sim 0.5$.
For $q=2$, we find similarly that the selected tail exponent 
remains the same within the whole delocalized phase
 and that it is of order $\mu_2^{selec} \sim 0.25$.
On Fig. \ref{fighistoc1c2} (b), we compare for $W=1$
the probability distributions $H_{q=1}(\ln d)$ and  $H_{q=2}(\ln d)$
of the variables $\ln d^{(q=1)}$  :
the exponents of Eq. \ref{Dtail} are respectively   $\mu_1^{selec} \sim 0.5$
and $\mu_2^{selec} \sim 0.25$.

From Eq. \ref{muselec}, we obtains that
 the Lyapunov exponents satisfy the identities
\begin{eqnarray}
\lambda_q^{selec} = q \lambda_1^{selec}
\label{selectionqidentity}
\end{eqnarray}
As explained in the text around Eq. \ref{LyapunovIqidentity}, this identity
is important to understand the decay of I.P.R.
 with the number $N$ of generations
in the delocalized phase.

\subsection{ Finite-size corrections introduced by the pool method }

\label{fspool}

As explained in section \ref{poolmethod}, the pool method consists in representing
a probability distribution $P(x)$ by a large
 number $M_{pool}$ of random variables $\{x_i\}$.
In particular this introduces a cut-off in the tail of $P(x)$ around $x_{max}$
with $P(x_{max}) \sim 1/M_{pool}$. In the field of traveling waves,
 the presence of such a cut-off has been much studied 
(see \cite{brunet} and references therein) : the leading correction to
the selected velocity is usually logarithmic in $M_{pool}$
\begin{eqnarray}
 v_{selec}(M_{pool}) -  v_{selec}(\infty)
 \sim \frac{J}{ ( \ln M_{pool})^{\alpha} }
\label{vpool}
\end{eqnarray}
For our present problem, where the critical point
 corresponds to a vanishing velocity $v=0$,
we thus expect that the true critical point $W_c(\infty)$ corresponding to $v_{selec}(\infty)=0$
will be shifted towards a pool-dependent pseudo-critical point $W_c(M_{pool})$
corresponding to $v_{selec}(M_{pool})=0$. As explained in the text,
all the traveling waves encountered in the present paper vanish linearly
in $(W-W_c)$. We thus expect that the difference
$[W_c(M_{pool})-W_c(\infty)]$ will also decay only
 logarithmically as $1/( \ln M_{pool})^{\alpha}$.
From the analysis presented in Appendix A and B, it is obvious that the pool dependent 
critical point for the traveling wave of the Landauer transmission (see Appendix A)
and for the auxiliary variable $C^{(1)}$ or $D^{(1)}$ (see Appendix B) are the same
because they are defined by the same condition in terms of the eigenvalue $\Lambda_0(\beta)$
of the integral kernel. However, the other variables $C^{(q)}$ or $D^{(q)}$
that have the same true critical point $W_c(\infty)$, 
will not have the same pool-dependent critical point $W_c(M_{pool})$, because the constant $J$ 
in Eq. \ref{vpool} will depend on $q$. This is indeed what we observe with our numerical
computations with the pool $M_{pool}=10^5$ : the pseudo-critical point $W_c(M_{pool})$
for the variable $C^{(2)}$ is below the pseudo-critical point $W_c(M_{pool})$
for the variable $C^{(1)}$. We observe that the difference between the two 
becomes smaller for the bigger pool $M_{pool}=10^6$.
In conclusion, numerical studies based on the pool method are 
valid for observables that are related to a single traveling wave,
but one cannot study the critical properties of observables
that depend on two distinct traveling waves that have different pseudo-critical points.
This is why in the text, 
we have not been able to present reliable numerical results for the statistics
of I.P.R. $I_2$ that depends on both $C^{(1)}$ and $C^{(2)}$.

\section{Reminder on the traveling/non-traveling phase transition 
studied in Ref. \cite{simon}}

In this Appendix, we briefly recall the finite-size scaling properties
of the traveling/non-traveling phase transition studied in Ref. \cite{simon},
because these properties are useful to interpret our numerical results described in the text.

For a one-dimensional branching random walk
in the presence of an absorbing wall moving at a constant velocity $v$,
the survival probability $Q(x,t)$ presents a phase transition at $v=v_v$
 with the following
critical behaviors (see more details in \cite{simon}) :

(i) for $v<v_c$, it converges exponentially in time towards a finite limit 
$Q^*(x)>0$
\begin{eqnarray}
Q(x,t) \simeq Q^*(x) + e^{- \frac{t}{\tau}} \phi(x)
\label{Qbelow}
\end{eqnarray}
where the limit $Q^*(x)$ presents an essential singularity
\begin{eqnarray}
 Q^*(x) \simeq e^{- \frac{cte}{(v_c-v)^{\kappa}} } \ \ {\rm with }
\ \  \kappa=\frac{1}{2}
\label{Qessential}
\end{eqnarray}
and where the relaxation time $\tau$ diverges as
\begin{eqnarray}
\tau \oppropto \frac{1}{(v_c-v)^{\nu} } \ \ {\rm with }
\ \  \nu=\frac{3}{2}
\label{Qtaurelax}
\end{eqnarray}

(ii) for $v>v_c$, it converges exponentially in time towards zero
\begin{eqnarray}
Q(x,t) \simeq e^{- \frac{t}{{\tilde \tau} }}
\label{Qabove}
\end{eqnarray}
where the relaxation time ${\tilde \tau}$ diverges as
\begin{eqnarray}
\tilde \tau \oppropto \frac{1}{ (v_c-v)^{ \tilde \nu} } \ \ {\rm with }
\ \  \tilde \nu=1
\label{Qtautyp}
\end{eqnarray}

(iii) exactly at criticality, it converges towards zero with a stretched exponential 
\begin{eqnarray}
Q(x,t) \simeq e^{- (cte) t^{\rho} }  \ \ {\rm with }
\ \  \rho=\frac{1}{3}
\label{Qcriti}
\end{eqnarray}

(iv) the critical region is described by some finite-size scaling form 
governed by the relaxation time $\tau$ that appears in Eq. \ref{Qbelow}
(and not by the relaxation time ${\tilde \tau}$  appearing in Eq. \ref{Qabove})
\begin{eqnarray}
\ln Q(x,t) \opsimeq - t^{\rho} G \left( t^{1/\nu} (v_c-v)  \right)
\label{Qfss}
\end{eqnarray}
where the scaling function $G(u)$ has the following asymptotic behaviors.
For $u \to +\infty$, has the one has the power-law
\begin{eqnarray}
G(u) \opsimeq_{u \to +\infty} u^{-\nu \rho}
\label{qfssplus}
\end{eqnarray}
to recover the finite limit of Eq. \ref{Qessential} and one has the scaling relation
\begin{eqnarray}
\kappa=\nu \rho
\label{kapparelation}
\end{eqnarray}
For $u \to -\infty$, one has the power-law
\begin{eqnarray}
G(u) \opsimeq_{u \to -\infty} (-u)^{\nu (1-\rho)}
\label{qfssmoins}
\end{eqnarray}
to recover the exponential decay of Eq. \ref{Qabove},
and one has the scaling relation 
\begin{eqnarray}
\tilde \nu= \nu (1-\rho)
\label{nutyprelation}
\end{eqnarray}
For the problem considered in \cite{simon} all the exponents are exactly known
$\tilde \nu=1$, $\nu=3/2$, $\rho=1/3$ and $\kappa=1/2$.

In our numerical results presented in the text
concerning the traveling/non-traveling phase transitions
that occur in Anderson localization on the Bethe lattice, we find very similar
critical behaviors : the velocity of the traveling wave vanishes linearly
with $\tilde \nu=1$, and the finite-size scaling in the critical region
is governed by the other exponent $\nu$ that appears in the relaxation towards
the finite value of the non-traveling phase.

\end{document}